\documentclass[%
 reprint,
 superscriptaddress,
 amsmath,amssymb,
 aps,
 pre,
]{revtex4-2}

\usepackage{graphicx}% Include figure files
\usepackage{dcolumn}% Align table columns on decimal point
\usepackage{bm}% bold math
\newcommand{\probP}{\text{I\kern-0.15em P}}
\usepackage{xcolor}
\usepackage{float}

\begin{document}

\preprint{APS/123-QED}

\title{Fractional Brownian motion with fluctuating diffusivities}% Force line breaks with \\

\author{Adrian Pacheco-Pozo}
    \affiliation{Department of Electrical and Computer Engineering and School of Biomedical Engineering, Colorado State University, Fort Collins, CO 80523, USA}
    
\author{Diego Krapf}
    \email{Corresponding author: diego.krapf@colostate.edu}
    \affiliation{Department of Electrical and Computer Engineering and School of Biomedical Engineering, Colorado State University, Fort Collins, CO 80523, USA}

\date{\today}% It is always \today, today,

\begin{abstract}
Despite the success of fractional Brownian motion (fBm) in modeling systems that exhibit anomalous diffusion due to temporal correlations, recent experimental and theoretical studies highlight the necessity for a more comprehensive approach of a generalization that incorporates heterogeneities in either the tracers or the environment. This work presents a modification of Levy's representation of fBm for the case in which the generalized diffusion coefficient is a stochastic process. We derive analytical expressions for the autocovariance function and both ensemble- and time-averaged mean squared displacements. Further, we validate the efficacy of the developed framework in two-state systems, comparing analytical asymptotic expressions with numerical simulations. 
\end{abstract}

\maketitle

%\tableofcontents

\section{Introduction}

% ---- Context ----

Anomalous diffusion processes are widespread in diverse disciplines, including nanoscale physics \cite{kindermann2017nonergodic}, cell and molecular biology \cite{weigel2011ergodic,barkai2012strange,manzo2015weak,krapf2019strange,sabri2020elucidating}, ecology  \cite{vilk2022unravelling,vilk2022ergodicity}, and finance \cite{bouchaud2005subtle,scalas2006application,plerou2000economic}. These processes are characterized by a non-linear time dependence of the mean square displacement (MSD), typically taking a power-law form $\langle X^2(t) \rangle \propto t^{\alpha}$ with an anomalous exponent $0<\alpha<2$. Here, the angular brackets denote averaging over an ensemble of trajectories. The diffusion is classified as subdiffusive when $0<\alpha<1$ and superdiffusive when $1<\alpha<2$. Normal diffusion is recovered in the limiting case $\alpha = 1$. Several mathematical models have been proposed to reproduce such  MSD \cite{hofling2013anomalous,metzler2014anomalous,manzo2015review,krapf2015mechanisms, shen2017single,munoz2021objective}. Among these models, fractional Brownian motion (fBm), a Gaussian process possessing temporal correlations \cite{kolmogorov1940wienersche,Mandelbrot1968}, has been widely used to model systems exhibiting anomalous diffusion with temporal correlations \cite{Szymanski2009,guigas2007probing,Magdziarz2009,jeon2011vivo,jeon2013anomalous,Sadegh2017}. 

Despite the success of fBm in modeling correlated random walks, experimental measurements often reveal marked heterogeneities in biological environments, highlighting the need for a generalization of fBm where its parameters change over time \cite{waigh2023heterogeneous,lanoiselee2018diffusion}. These complexities are usually due to fluctuations in the tracer particles or the medium where the diffusion takes place. Examples of such heterogeneous systems include the diffusion of proteins and lipids in the plasma membrane \cite{He2016,Jeon2016,Sikora2017,Weron2017}, intracellular transport of endosomes and lysosomes \cite{Han2020,Fedotov2023}, and DNA-binding proteins \cite{Loverdo2009}, among others \cite{Munoz2023,Bronstein2009,Sabri2020}. 

From a theoretical approach, heterogeneities have been modeled extensively as Brownian particles with fluctuating diffusivity. Examples of this type of motion include diffusing diffusivities \cite{chubynsky2014diffusing,chechkin2017brownian} and the annealed transit time model \cite{massignan2014nonergodic}. To address temporal correlations in heterogeneous systems, several stochastic processes have been proposed as modifications of fBm to model heterogeneous transport \cite{Wang2020,Wang2020_1,Fox2021,Szarek2022,Balcerek2022,Balcerek2023,Wang2023,Slezak2023}. 
Among them, the diffusion of particles stochastically switching between two states has been studied using numerical simulations \cite{Balcerek2023}. In this previous work, each state $i$ was characterized by a generalized diffusion coefficient $D_i$, a Hurst exponent $H_i$, and independent and identical distributed (i.i.d) dwell times. To maintain time correlations in the switching fBm (sfBm), a modification of the integral representation of the fBm process was used \cite{Levy1953,Wang2023}. The sfBm is of particular interest when modeling biological systems, such as the dynamics of nano-scale particles in the cytoplasm of mammalian cells \cite{Sabri2020,janczura2021identifying}. Numerical simulations revealed asymptotic scalings of the temporal average MSD and the power spectral density. However, an analytical framework to study processes having rich dynamics, such as switching between states while keeping their temporal correlations is still missing.

%From a theoretical approach, heterogeneities have been modeled extensively as Brownian particles with fluctuating diffusivity \cite{Miyaguchi2016}. Examples of this type of motion include diffusing diffusivities \cite{chubynsky2014diffusing,chechkin2017brownian} and the annealed transit time model \cite{massignan2014nonergodic}. To address temporal correlations in heterogeneous systems, several stochastic processes have been proposed as modifications of Mandelbrot's fBm to model heterogeneous transport \cite{Wang2020,Wang2020_1,Fox2021}. Lately, several works \cite{Szarek2022,Balcerek2022,Balcerek2023,Wang2023,Slezak2023} have consider Levy's formulation of fBm \cite{Levy1953} since both its formulation and numerical implementation is simpler than that of Mandelbrot. Among these works, Belcerek and colaborators \cite{Balcerek2023} used Levy's fBm to numerically studied the diffusion of particles stochastically switching between two states (dichotomous process). This type of switching fBm (sfBm) is of particular interest when modeling biological systems, such as the dynamics of nano-scale particles in the cytoplasm of mammalian cells \cite{Sabri2020,janczura2021identifying}. Their simulations revealed asymptotic scalings of the temporal average MSD and the power spectral density. However, a simple, analytical framework to study processes having rich dynamics, such as switching between states while keeping their temporal correlations is still missing.

In this work, we present a framework for addressing systems exhibiting temporal correlations akin to fBm, while encompassing rich dynamics characterized by a fluctuating generalized diffusion coefficient. The presented model is valid for the diffusion coefficient being any (non-negative) stochastic process. Importantly, temporal correlations are maintained throughout the whole trajectory. The powerful yet simple applicability of the proposed framework is applied to two-state systems for which asymptotic expressions can be compared to numerical simulations. This model is closely related to the switching fBM ~\cite{Balcerek2023} and can be seen as a generalization of the uncorrelated dichotomous model studied in Ref.~\cite{Miyaguchi2016}.
We focus on dichotomous processes with dwell times that have either an exponential or a heavy-tailed distribution, which are both relevant to diffusion in complex systems. Exponential distributions of switching times have been observed for the motion of inert particles in the cytoplasm of live cells \cite{Sabri2020,janczura2021identifying,dieball2022scattering}. Heavy-tailed power-law distributions in the dwell times yield aging and non-ergodicity, and they have been observed for protein dynamics in the plasma membrane \cite{Weron2017,weigel2013quantifying,manzo2015weak}, intracellular transport of insulin granules \cite{tabei2013intracellular}, and the internal dynamics in globular proteins \cite{hu2016dynamics,yang2003protein,krapf2019strange}.

This article is structured as follows: In Sec.~\ref{sec:gen_frame} we derive the general framework that will be used in the context of two-state systems. In Sec.~\ref{sec:two_state} we introduce two-state systems and derive, using our framework, the corresponding asymptotic expressions for the MSD, which we then compare with numerical simulations for each particular case. Sec.~\ref{sec:summary} presents a summary and concluding remarks.

% ---- Section ----

\section{Framework for $D(t)$ being a stochastic process \label{sec:gen_frame}}
Mandelbrot's fBm $B_H(t)$ is a zero-mean, continuous Gaussian process, characterized by a Hurst exponent ${H \in (0,1)}$ related to the anomalous exponent by ${H=\alpha /2}$ \cite{Mandelbrot1968}. It is defined by an autocovariance  
\begin{equation}
\langle B_H(t_1)B_H(t_2) \rangle = D (t_1^{2H}+t_2^{2H}-|t_1 - t_2|^{2H}),
\label{eq:auto_fbm}
\end{equation}
where $D$ is the generalized diffusion coefficient with units of $\mathrm{length}^2/\mathrm{time}^{2H}$. From Eq.~(\ref{eq:auto_fbm}) follows that the MSD exhibits anomalous diffusion of the form
\begin{equation}
\langle B_H^2(t) \rangle = 2 D t^{2H}.
\label{eq:MSD_fbm}
\end{equation}
The motion is then classified according to the value of $H$ as subdiffusion when ${0<H<1/2}$ and as superdiffusion when ${1/2<H<1}$. Standard Brownian motion is recovered for ${H = 1/2}$. An alternative form of fBm consists of L\'evy's non-equilibrated formulation \cite{levy1953random}, which is written in terms of the Riemann-Liouville fractional integral as
\begin{equation}
B_H(t) =  \sqrt{4HD} \int_0^t (t - \tau)^{H - 1/2} \; \xi(\tau) d\tau,
\label{eq:fBm_levy}
\end{equation}
where $\xi(t)$ is zero-mean Gaussian white noise with delta-correlations, that is, $\langle \xi(t) \rangle = 0$, and $\langle \xi(t_1)\xi(t_2) \rangle = \delta(t_2 - t_1)$.

Following recent works \cite{Balcerek2023,Wang2023}, we consider fractional Brownian motion with fluctuating diffusivity as the process in Eq.~(\ref{eq:fBm_levy}) with a generalized diffusion coefficient being a stochastic process $D(t)$,
\begin{equation}
X(t) =  \sqrt{4H} \int_0^t \sqrt{D(\tau)} \; (t - \tau)^{H - 1/2} \; \xi(\tau) d\tau.
\label{eq:process}
\end{equation}
This process has two sources of randomness: one due to variations of the Brownian motion or Gaussian white noise $\xi(t)$, and another due to the generalized diffusion coefficient $D(t)$ being a stochastic process. Thus, when computing the mean of an observable $Q$, denoted as $\langle Q \rangle$, two averages must be taken, one over the Gaussian noise, denoted as $\langle \cdots \rangle_{\xi}$, and one over the different realizations of the generalized diffusion coefficient $D(t)$, denoted as $\langle \cdots \rangle_{D}$, therefore $\langle Q \rangle = \langle \langle Q \rangle_{\xi} \rangle_{D}$. We will refer to the first average as the average over the noise, whereas the second average will be denoted as an average over the disorder.

% ---- Subsection ----

\subsection{Mean square displacement}

To find the MSD $\langle X^2(t) \rangle$, we first average over the noise,
\begin{multline}
\langle X^2(t) \rangle_{\xi} = 4H \int_0^t dt_1 \int_0^t dt_2 \sqrt{D(t_1) D(t_2)} \; \times \\
\times (t - t_1)^{H - 1/2} (t - t_2)^{H - 1/2} \langle \xi(t_1)\xi(t_2) \rangle_{\xi}.
\label{eq:m1}
\end{multline}
Because the white noise $\xi(t)$ is delta-correlated, this expression simplifies to
\begin{equation}
\langle X^2(t) \rangle_{\xi} = 4H \int_0^t  D(t_1) (t - t_1)^{2H - 1} dt_1. 
\label{eq:m2}
\end{equation}
Next, we average over the disorder, obtaining
\begin{equation}
\langle \langle X^2(t) \rangle_{\xi} \rangle_{D} = 4 H \int_0^t  \langle D(t_1) \rangle_{D} \; (t - t_1)^{2H - 1} dt_1,
\label{eq:MSD}
\end{equation}
which has the form of a convolution, with a Laplace transform that can be written as
\begin{equation}
\langle X^2(s)\rangle = \frac{2 \; \Gamma(2H + 1)}{s^{2H}} \langle D(s) \rangle_D,
\label{eq:MSD_laplace}
\end{equation}
where $\Gamma(x)$ is the Gamma-function. Thus, by finding the Laplace transform of the mean of the generalized diffusion coefficient process $\langle D(s) \rangle$, the MSD can be readily obtained. In what follows, we will drop for convenience the subscript $D$.

When the generalized diffusion coefficient $D(t)$ is a stationary process, its mean is time-independent ($\langle D_{\text{st}}(t) \rangle = \langle D_{\text{st}} \rangle$) and can be taken out of the integral in Eq.~(\ref{eq:MSD}). So, the MSD reduces to
\begin{equation}
\langle X_{\text{st}}^2(t) \rangle = 2 \langle D_{\text{st}} \rangle t^{2H},
\label{eq:MSD_stat}
\end{equation}
which has the well-known form of Eq.~(\ref{eq:MSD_fbm}) with effective diffusivity $\langle D_{\text{st}} \rangle$.

% ---- Subsection ----

\subsection{Autocovariance}
The covariance function can be expressed as
\begin{multline}
\langle X(t_1)X(t_2) \rangle = \frac{1}{2} (\langle X^2(t_1) \rangle + \\ + \langle X^2(t_2) \rangle - \langle [X(t_2) - X(t_1)]^2 \rangle).
\label{eq:cov}
\end{multline}
The first two terms in the r.h.s. of this expression are the MSD (Eq.~(\ref{eq:MSD})) at times $t_1$ and $t_2$. The last term corresponds to the mean of the squared increments, which can be found using a similar procedure as that of Ref.~\cite{Marinucci1999}. Assuming, without loss of generality, that $t_2 > t_1$, we write the increments as
\begin{multline}
X(t_2) - X(t_1) =  \int_{t_1}^{t_2} \sqrt{4H D(u)} \times \\
\times (t_2 - u)^{H-1/2} \xi(u) du + \int_0^{t_1} \sqrt{4HD(v) } \times \\
\times \left[( t_2 - v)^{H - 1/2} - ( t_1 - v)^{H-1/2} \right] \xi(v) dv.
\end{multline}
Then, taking the square of this quantity and averaging it first over the noise and then over the disorder as in Eqs.~(\ref{eq:m1})-(\ref{eq:MSD}), we can write the mean of the squared increments as
\begin{multline}
\langle [X(t_2) - X(t_1)]^2 \rangle =\\
= 4H\int_{t_1}^{t_2} \langle D(u) \rangle (t_2 - u)^{2H-1} du + \\
 + 4H \int_0^{t_1} \langle D(v) \rangle \left[( t_2 - v)^{H - 1/2} - ( t_1 - v)^{H-1/2} \right]^2 dv.
\end{multline}
Let us now introduce the variable $\tau = t_2 - t_1$ and make the change of variables $w = (t_1 - v) / \tau$ which allow us to rewrite the last equation as
\begin{multline}
\langle [X(t_2) - X(t_1)]^2 \rangle =\\
= 4H \int_{t_1}^{t_2} \langle D(u) \rangle (t_2 - u)^{2H-1} du + 4H \tau^{2H} \times \\
\times \int_0^{t_1 / \tau} \langle D(t_1-\tau w) \rangle\left[( 1 + w )^{H - 1/2} - w^{H-1/2} \right]^2 dw.
\end{multline}
Next, working in the limit $t_1 / \tau \to 0$, the second integral vanishes and this last expression takes the form
\begin{multline}
\langle [X(t_1) - X(t_2)]^2 \rangle = \\
= 4H \int_{t_1}^{t_2} \langle D(u) \rangle (t_2 - u)^{2H-1} du,
\label{eq:increments}
\end{multline}
which depends on the form of $\langle D(t) \rangle$. This limit ensures that fBm as defined in Eq.~(\ref{eq:process}) with constant generalized diffusion coefficient has the same properties as the Mandelbrot's fBm \cite{Mandelbrot1968}. For the case where the process $D(t)$ is stationary, this last integral can be solved exactly, and, for any $t_1$ and $t_2$, it reads
\begin{equation}
\langle [X_{\text{st}}(t_1) - X_{\text{st}}(t_2)]^2 \rangle = 2 \langle D_{\text{st}} \rangle |t_1 -t_2|^{2H}.
\label{eq:stat_inc}
\end{equation}
Thus, the covariance function (Eq.~(\ref{eq:cov})) is
\begin{equation}
\langle X_{\text{st}}(t_1)X_{\text{st}}(t_2) \rangle = \langle D_{\text{st}} \rangle \left[ t_1^{2H} + t_2^{2H} - |t_1 -t_2|^{2H} \right],
\end{equation}
which has the form of Eq.~(\ref{eq:auto_fbm}). From this last equation and Eq.~(\ref{eq:MSD_stat}), it is clear that, when the process $X(t)$ has a wide-sense stationary generalized diffusion coefficient $D(t)$, it behaves as a standard fBm with an effective generalized diffusion coefficient  $\langle D_{\text{st}} \rangle$.

\subsection{Temporal average MSD}

A quantity that is widely used in single particle tracking analysis is the temporal average MSD (TAMSD), defined for an individual trajectory as
\begin{equation}
\overline{\delta^2(\Delta)} = \frac{1}{T-\Delta} \int_0^{T-\Delta} [X(t+\Delta) - X(t)]^2 dt.
\end{equation}
Using Eq.~(\ref{eq:cov}), for $T \gg \Delta$, the ensemble average of the TAMSD can be written as
\begin{multline}
\langle \overline{\delta^2(\Delta)} \rangle \approx \frac{1}{T} \int_0^{T} \left[ \langle X^2(t+\Delta) \rangle + \langle X^2(t) \rangle \right. - \\
- \left. 2\langle X(t+\Delta)X(t)\rangle \right] dt.
\end{multline}
Henceforth, we will refer to $\langle \overline{\delta^2(\Delta)} \rangle$ as the TAMSD. Taking the Laplace transform in the variable $T$ on both sides,
\begin{multline}
\mathcal{L}_T\left[T\langle \overline{\delta^2(\Delta)} \rangle \right] = \frac{1}{s} \mathcal{L}_t\left[ \langle X^2(t+\Delta) \rangle + \langle X^2(t) \rangle - \right. \\
- \left. 2 \langle X(t+\Delta)X(t)\rangle \right],
\label{eq:laplace_T}
\end{multline}
where we have introduced a subscript on the Laplace transforms to remove any possible ambiguity as to which variable is being used, $T$ or $t$. The first term in the right-hand side of Eq.~(\ref{eq:laplace_T}) corresponds to the Laplace transform of a ``shifted'' MSD, whereas the second one corresponds to the Laplace transform of the MSD. This expression, then, takes the form
\begin{multline}
\mathcal{L}_T\left[T\langle \overline{\delta^2(\Delta)} \rangle \right] = \frac{1}{s} \left[ (e^{\Delta s}+1) \langle X^2(s) \rangle  - \right. \\
\left. - 2 \mathcal{L}_t( \langle X(t+\Delta)X(t)\rangle) \right].
\label{eq:laplace_TAMSD_calcu}
\end{multline}
The exact Laplace transform for the covariance function involves a modified Bessel function of the second kind and is presented in Appendix~\ref{app:LT_covariance}. Here, we are mostly interested in the large time asymptotics, which correspond to the small $s$ behavior. In this regime, the Laplace transform has the form
\begin{multline}
\mathcal{L}_t\left(\langle X(t+\Delta)X(t)\rangle \right) = 2 \langle D(s) \rangle \frac{\Gamma(2H+1)}{s^{2H}} \times \\
\times \left[ 1 + \frac{\Delta s}{2} -  \frac{\Gamma(1-H)}{\Gamma(H+1)} \left( \frac{\Delta s}{4} \right)^{2H} + O(s^{2H+1}) \right],
\end{multline}
where $O(s^{n})$ is Landau's big-O notation.
By substituting this result into Eq.~(\ref{eq:laplace_TAMSD_calcu}) and taking into account that $e^{\Delta s} = 1 + \Delta s + O(s^2)$, the final expression reads
\begin{equation}
\mathcal{L}_T\left[T\langle \overline{\delta^2(\Delta)} \rangle \right] = \frac{\langle D(s) \rangle}{s} \frac{\Gamma(2H+1)  \Gamma(1-H)}{4^{2H - 1} \Gamma(H+1)} \Delta^{2H}.
\label{eq:TAMSD_laplace}
\end{equation}
Finally, taking the inverse Laplace transform we obtain the TAMSD in time domain,
\begin{equation}
\langle \overline{\delta^2(\Delta)} \rangle = \frac{1}{T} \mathcal{L}_T^{-1} \left[ \frac{\langle D(s) \rangle}{s} \right] \frac{\Gamma(2H+1)  \Gamma(1-H)}{4^{2H - 1} \Gamma(H+1)} \Delta^{2H},
\label{eq:TAMSD}
\end{equation}
for $T \gg \Delta$. For $H = 1/2$, the result for Brownian motion with fluctuating diffusivity is recovered \cite{Miyaguchi2016}. Moreover, when the process $D(t)$ is wide-sense stationary, one has
\begin{equation}
\langle D_{\text{st}}(s) \rangle = \frac{\langle D_{\text{st}} \rangle}{s},
\end{equation}
which is the Laplace transform of a constant. Then, under the wide-sense stationarity condition, Eq.~(\ref{eq:TAMSD}) reduces to
\begin{equation}
\langle \overline{\delta^2(\Delta)} \rangle =  \frac{\Gamma(2H+1) \Gamma(1-H)}{4^{2H - 1} \Gamma(H+1)} \langle D_{\text{st}} \rangle \Delta^{2H},
\label{eq:TAMSD_stat}
\end{equation}
which by comparison with Eq.~(\ref{eq:MSD_stat}), one concludes that the system exhibits ultraweak ergodicity breaking, that is, the time- and ensemble-averaged MSDs have the same scaling but the prefactor is different. 
The prefactor in the TAMSD approaches $4$ as $H$ approaches zero (i.e. double the prefactor of the ensemble-averaged MSD) and it diverges as $H$ approaches unity, the ballistic case. In the whole range $0\le H\le1$, this prefactor is bounded below by $2$, reaching this minimum when $H=1/2$ (standard Brownian motion). Thus, when $H=1/2$  the ergodic behavior is recovered.

% ---- Section ----

\section{Two-state system with heavy-tailed sojourn times \label{sec:two_state}}

To illustrate the advantages of our approach, we consider a two-state system. In this model, the particle can be in the states `$+$' or `$-$'. These states are characterized by generalized diffusion coefficients, $D^+$ and $D^-$, and dwell time distributions, $\psi_+(t)$ and $\psi_-(t)$. Both states are considered to have the same Hurst exponent $H$. This class of switching fBm models has been analyzed in Ref.~\cite{Balcerek2023} using numerical simulations and is related to the Brownian with fluctuating diffusivity studied in Ref.~\cite{Miyaguchi2016}. 

We consider dwell times that have either a heavy-tailed distribution with infinite mean or that are exponentially distributed. Typical heavy-tailed distributions are asymptotically described by a power-law of the form
\begin{equation}
\psi_{\text{PL}}(t) \sim \frac{a}{|\Gamma(-\alpha)| t^{1+\alpha}},
\label{eq:power-law_time}
\end{equation}
where $0<\alpha<1$, and $a$ is a constant. The condition $\alpha <1$ leads to $t$ not having a first moment. The Laplace transform of $\psi_{\text{PL}}(t)$, i.e., the moment generating function (MGF) of the dwell times, is
\begin{equation}
\psi_{\text{PL}}(s) = 1 - a s^{\alpha} + O(s).
\label{eq:lap_zero_one}
\end{equation}

When $\psi_{\pm}$ are exponential distributions, all the moments exist and we have
\begin{equation}
\psi_{\text{exp}}(t) = \frac{1}{\tau} \exp \left( - t/\tau \right),
\end{equation}
with $\tau$ being the first moment. The Laplace transform of this distribution (MGF of the dwell times) is
\begin{equation}
\psi_{\text{exp}}(s) = \frac{1}{1+ \tau s}  = 1 - \tau s + O(s^2),
\end{equation}
where the right-hand side is useful for small $s$ (large $t$) approximations.

When at least one of the two distributions $\psi_{\pm}(t)$ has a heavy-tailed form with infinite mean, the two-state system does not reach a steady state \cite{Miyaguchi2016}. Nevertheless, regardless of the distributions being of power-law or exponential form, the first moment of the process $D(t)$ in Laplace domain has the general small $s$ asymptotic \cite{Miyaguchi2016},
\begin{equation}
\langle D(s) \rangle  \approx  \frac{A}{s} + B \frac{\Gamma(\beta)}{s^{\beta}},
\label{eq:neq_meanK_Laplace}
\end{equation}
where $A, B$ and $\beta$ are constants that depend on $\psi_{\pm}(s)$. Taking the inverse Laplace transform  of this last expression one finds
\begin{equation}
\langle D(t) \rangle \approx A + B t^{\beta-1},
\label{eq:neq_meanK}
\end{equation}
for large $t$. Thus, by computing either $\langle D(s) \rangle$ at small $s$ or $\langle D(t) \rangle$ at large $t$, the constants $A$, $B$, and $\beta$ are found and, in turn, the statistics of the process can be readily derived as shown below. 

The asymptotic behavior of the MSD can be found by substituting Eq.~(\ref{eq:neq_meanK_Laplace}) into Eq.~(\ref{eq:MSD_laplace}) and taking the inverse Laplace transform,
\begin{equation}
\langle X^2(t) \rangle \approx 2 A t^{2H} + \frac{2 B \Gamma(\beta) \Gamma(2H + 1)}{\Gamma(\beta + 2H)} t^{\beta + 2 H - 1},
\label{eq:asym_MSD}
\end{equation}
for large $t$. This expression is valid for any value $0<H<1$. Moreover, for normal diffusion ($H = 1/2$), Eq.~(\ref{eq:asym_MSD}) reduces to
\begin{equation}
\langle X^2(t) \rangle \approx 2 A t + \frac{2 B}{\beta} t^{\beta},
\end{equation}
which agrees with the result obtained in Ref.~\cite{Miyaguchi2016}.

To find the TAMSD, we substitute Eq.~(\ref{eq:neq_meanK_Laplace}) into Eq.~(\ref{eq:TAMSD}) and invert the Laplace transform, 
\begin{equation}
\langle \overline{\delta^2(\Delta)} \rangle = \left( A + \frac{B}{\beta} T^{\beta-1} \right) \frac{\Gamma(2H+1)  \Gamma(1-H)}{4^{2H - 1} \Gamma(H+1)} \Delta^{2H},
\label{eq:asym_TAMSD}
\end{equation}
which is also valid for any $H$. Comparing the time- and ensemble-averaged MSD (Eqs.~(\ref{eq:asym_MSD}) and~(\ref{eq:asym_TAMSD})) also shows that the process displays ergodicity breaking.

\subsection*{Examples}
\begin{figure}[htp]
    \centering
    \includegraphics[width=0.9\columnwidth]{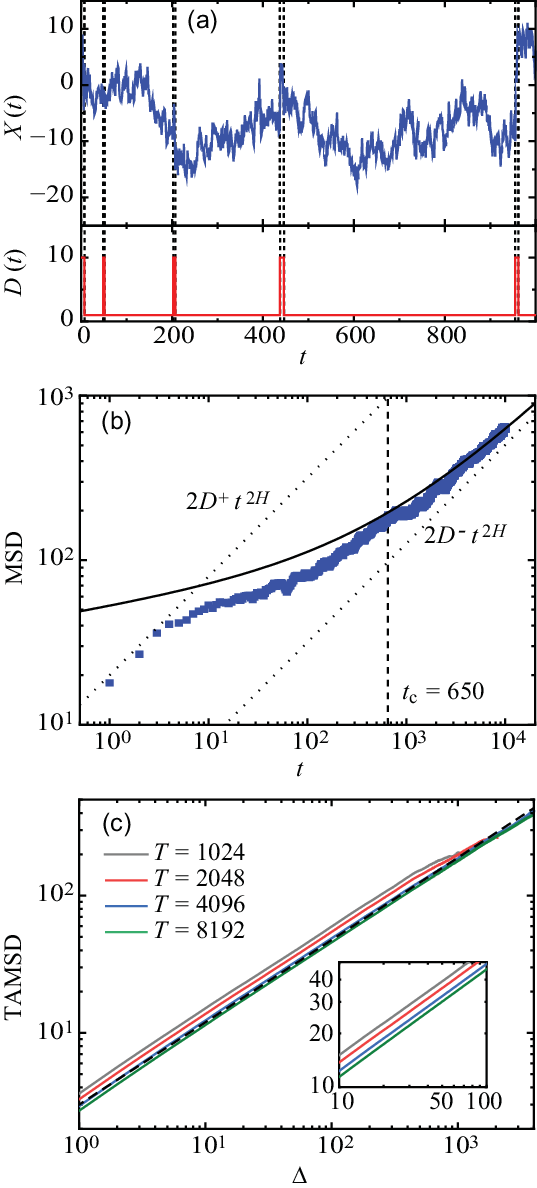}
    \caption{Case 1: Two-state system where both states have heavy-tailed dwell time distributions. The parameters in this example are $D^+ = 10$, $D^- = 1$, $\alpha_+ = 0.75$, $\alpha_- = 0.25$, $t_{0,+} = 3$, and $t_{0,-} = 25$. (a) Representative trajectory $X(t)$ together with $D(t)$. (b) The MSD is shown for numerical simulations as blue squares. The solid line shows the analytical asymptotic behavior given by Eq~(\ref{eq:MSD_c1}) and the dotted lines show the MSD of each of the two states. (c) TAMSD for different realization times, the dashed line represents the analytical asymptotic behavior given by Eq.~(\ref{eq:TAMSD_c1}) for $T = 8192$. The inset shows a close-up to highlight the dependence of TAMSD on the realization time $T$.
    }
    \label{fig:MSD_1}
\end{figure}
We now consider three cases to test the fractional Brownian motion with fluctuating diffusivity in two-state systems. Namely, (1) both dwell time distributions are heavy-tailed with infinite mean, (2) one of the dwell time distributions has infinite mean and the second is exponential, and (3) both distributions are exponential. We assume that $D^+ > D^- >0$. In order to compare the analytical results with numerical simulations, for asymptotic power law distributions we employ a Pareto distribution,
\begin{equation}
\psi_{\text{PL}}(t) = \frac{\alpha t_0^{\alpha}}{t^{\alpha+1}}, \quad \text{for } t \geq t_0,
\end{equation}
which has an exact Laplace transform (MGF of the dwell times)
\begin{equation}
\psi_{\text{PL}}(s) = \alpha t_0^{\alpha} s^{\alpha} \Gamma(-\alpha, t_0 s),
\end{equation}
with $\Gamma(x,y)$ being the upper incomplete Gamma function. In the limit $s\to 0$, this expression takes the form of Eq.~(\ref{eq:lap_zero_one})  with $a = \Gamma(1-\alpha) t_0^{\alpha}$.

% ---- Subsection ----

\subsection{Case 1: Diverging mean dwell times, $0 < \alpha_{\pm} <1$}

For $\alpha_{\pm} \in (0,1)$, with $\alpha_+ > \alpha_-$, the asymptotic behavior of the mean generalized diffusion coefficient takes the form \cite{Miyaguchi2016}
\begin{equation}
\langle D(t) \rangle \approx D^- + \frac{a_+}{a_-} \frac{(D^+ - D^-) t^{\alpha_--\alpha_+}}{\Gamma(\alpha_- - \alpha_+ + 1)},
\label{eq:case1_K}
\end{equation}
which converges to $D^-$ at large times. 
%$t\gg (a_-/a_+)^{1/(\alpha_- - \alpha_+)}$.
By comparison with Eq~(\ref{eq:neq_meanK}), one finds 
\begin{equation}
A = D^-, \;  B =\frac{a_+}{a_-} \frac{(D^+ - D^-)}{\Gamma(\beta)}, \; \text{and} \;  \beta = \alpha_--\alpha_+ + 1.
\end{equation}
Therefore, the asymptotic behavior of the MSD is 
\begin{multline}
\langle X^2(t)\rangle \approx 2 D^- t^{2H} +\\
+ 2 \frac{a_+}{a_-} \frac{\Gamma(2H+1)(D^+ - D^-)}{\Gamma(\alpha_- -\alpha_+ + 2H + 1)}  t^{\alpha_- - \alpha_+ + 2H},
\label{eq:MSD_c1}
\end{multline}
which converges to $2 D^- t^{2H}$ at large $t$. 

In order to evaluate this type of two-state systems, we simulated $1000$ trajectories of fractional Brownian motion with fluctuating diffusivity (see Appendix~\ref{app:Num}) with $H = 0.3$, $D^+ = 10$, $D^- = 1$, $\alpha_+ = 0.75$, $\alpha_- = 0.25$, $t_{0,+} = 3$, and $t_{0,-} = 25$. All the realizations start in the `$+$' state. Figure~\ref{fig:MSD_1}(a) shows one trajectory $X(t)$ alongside the process $D(t)$ used to compute it. Figure~\ref{fig:MSD_1}(b) shows the analytical MSD together with the numerical simulation results. The asymptotic behavior given by Eq.(\ref{eq:MSD_c1}) presents a crossover between two regimes with $\langle X^2(t)\rangle \sim t^{2H}$ and $\langle X^2(t)\rangle \sim t^{\alpha_+ - \alpha_- + 2H}$. This crossover takes place at the critical time 
\begin{equation}
t_\text{c} = \left[ \frac{a_+}{a_-} \frac{\Gamma(2H+1)}{\Gamma(\alpha_- -\alpha_+ + 2H + 1)} \frac{D^+ - D^- }{D^-} \right]^\frac{1}{\alpha_+ - \alpha_-},
\end{equation}
which, here, is $t_\text{c} = 650$.
The MSD of numerical simulations shown in Fig.~\ref{fig:MSD_1}(b) exhibits three different regimes. An initial regime where the MSD depends on the initial condition, i.e. the starting state, an intermediate regime, up to a time of the order of $t_c$ where the MSD has an anomalous exponent smaller than $2H$ and a long time asymptotic with the anomalous exponent being $2H$.

\begin{figure}[htp]
    \centering
    \includegraphics[width=0.9\columnwidth]{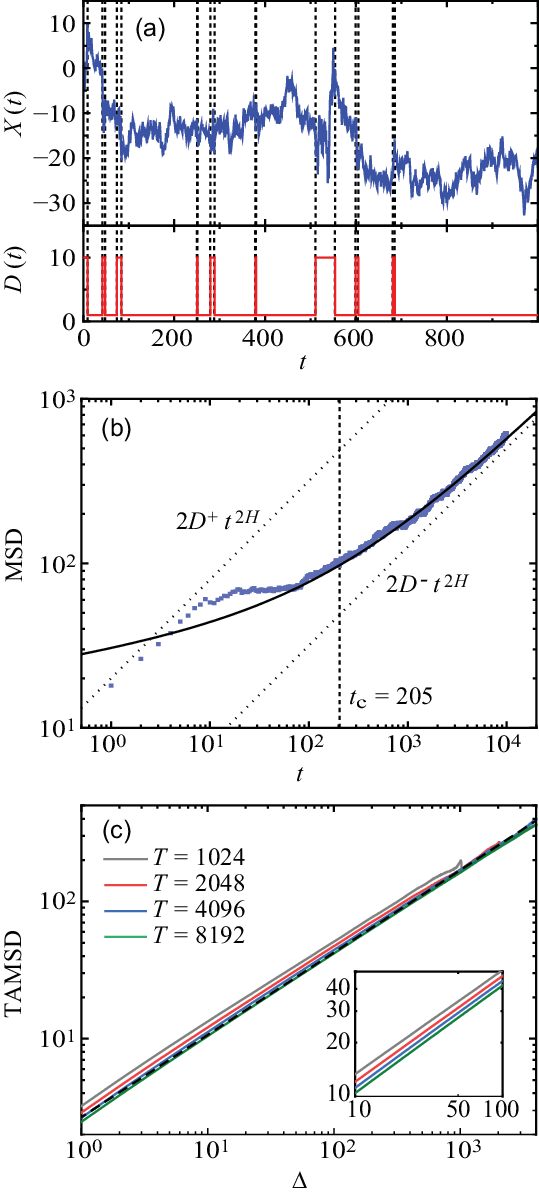}
    \caption{ Case 2: Two-state system where one state has an exponential distribution and the other state has a heavy-tailed dwell time distribution. The parameters in this example are $D^+ = 10$, $D^- = 1$, $\tau_{+} = 15$, $\alpha_- = 0.5$, and $t_{0,-} = 25$. (a) Representative trajectory $X(t)$ together with $D(t)$. (b) The MSD is shown for numerical simulations as blue squares. The solid line shows the analytical asymptotic behavior given by Eq~(\ref{eq:MSD_c1}) and the dotted lines show the MSD of each of the two states. (c) TAMSD for different realization times, the dashed line represents the analytical asymptotic behavior given by Eq.~(\ref{eq:TAMSD_c1}) for $T = 8192$. The inset shows a close-up to highlight the dependence of TAMSD on the realization time.
    }
    \label{fig:MSD_2}
\end{figure}

The TAMSD (Fig.~\ref{fig:MSD_1}(c)) for this case is
\begin{multline}
\langle \overline{\delta^2(\Delta)}\rangle =  \left[ D^- + \frac{a_+}{a_-} \frac{(D^+ - D^-)}{\Gamma(\alpha_- - \alpha_+ + 2)} T^{\alpha_- -\alpha_+} \right]  \times \\
\times \frac{\Gamma(2H+1)  \Gamma(1-H)}{4^{2H - 1} \Gamma(H+1)} \Delta^{2H}. 
\label{eq:TAMSD_c1}
\end{multline}
The TAMSD exhibits a dependence on the experimental time via a scaling $T^{\alpha_- -\alpha_+}$. In the long time limit, it converges to $\Gamma(2H+1)\Gamma(1-H) / [4^{2H - 1} \Gamma(H+1)] D^- \Delta^{2H}$. The large-$t$ aymptotic of the TAMSD, thus, differs from the ensemble-averaged MSD ($2 D^- \Delta^{2H}$) by a constant factor, showing ultraweak ergodicity breaking. 

% ---- Subsection ----

\subsection{Case 2: Exponential distribution $\psi_+(t)$ and heavy-tailed $\psi_-$, $0 < \alpha_{-}<1$}

For $\psi_+(t)$ being exponential and $\psi_-(t)$ being heavy-tailed with $\alpha_{-} \in (0,1)$, the asymptotic behavior of the mean generalized diffusion coefficient reads
\begin{equation}
\langle D(t) \rangle \approx D^- + \frac{\tau_+}{a_-} \frac{(D^+ - D^-)}{ \Gamma(\alpha_{-})} t^{\alpha_--1}.
\label{eq:case2_K}
\end{equation}
A comparison with Eq.~(\ref{eq:neq_meanK}) shows
\begin{equation}
A = D^-, \; B = \frac{\tau_+}{a_-}  \frac{(D^+ - D^-)}{a_+ \Gamma(\alpha_{-})}, \; \text{and} \; \beta = \alpha_-.
\end{equation}
Thus, the asymptotic behavior of the MSD is
\begin{multline}
\langle X^2(t)\rangle \approx 2 D^- t^{2H} +\\
+ 2 \frac{\tau_+}{a_-} \frac{\Gamma(2H+1) (D^+ - D^-)}{\Gamma(\alpha_- + 2H)}  t^{\alpha_- + 2H -1},
\label{eq:MSD_c2}
\end{multline}
which converges to $2 D^- t^{2H}$ at long times. 

We generated $1000$ realizations of the process starting in the `$+$' state, with  $H = 0.3$, $D^+ = 10$, $D^- = 1$, $\alpha_- = 0.5$, $\tau_{+} = 15$, and $t_{0,-} = 25$. Figure~\ref{fig:MSD_2}(a) shows a trajectory $X(t)$ alongside the process $D(t)$. Figure~\ref{fig:MSD_2}(b) shows a comparison between the asymptotic MSD and the one obtained from numerical simulations. In a similar way as for case 1, there is a crossover between diffusion regimes, with a crossover-time
\begin{equation}
t_\text{c} = \left[ \frac{\tau_+}{a_-} \frac{\Gamma(2H+1)}{\Gamma(\alpha_- + 2H)} \frac{D^+ - D^-}{D^-} \right]^{1 / (1-\alpha_+)},
\end{equation}
which, here, is $t_\text{c} = 205$.

The TAMSD [shown in Fig.~\ref{fig:MSD_2}(c)] 
\begin{multline}
\langle \overline{\delta^2(\Delta)}\rangle =  \left[ D^- + \frac{\tau_+}{a_-} \frac{(D^+ - D^-)}{ \Gamma(\alpha_{-}+1)} T^{\alpha_--1} \right]  \times \\
\times \frac{\Gamma(2H+1)  \Gamma(1-H)}{4^{2H - 1} \Gamma(H+1)} \Delta^{2H},
\label{eq:TAMSD_c3}
\end{multline}
which, again, in the long time limit, differs from the ensemble-averaged MSD ($2 D^- \Delta^{2H}$) by a constant factor (ultraweak non-ergodicity).

% ---- Subsection ----

\begin{figure}[htp]
    \centering
    \includegraphics[width=0.9\columnwidth]{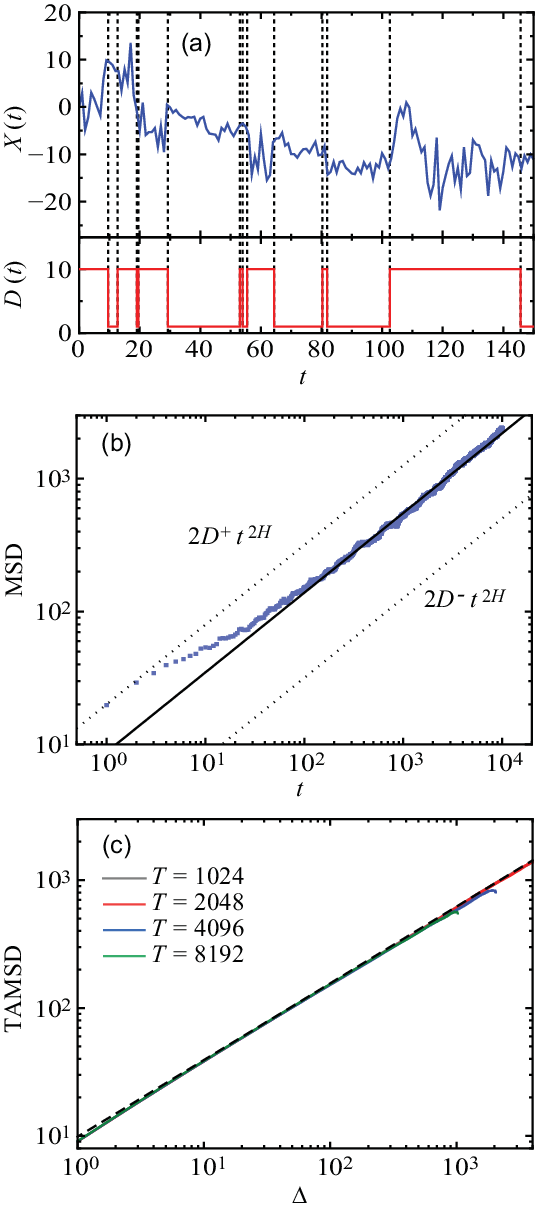}
    \caption{Case 3: Two-state system where both states have exponential dwell time distributions. The parameters in this example are $D^+ = 10$, $D^- = 1$, $\tau_{+} = 15$, and $\tau_{-} = 25$. (a) Representative trajectory $X(t)$ together with $D(t)$. (b) The MSD is shown for numerical simulations as blue squares. The solid line shows the analytical asymptotic behavior given by Eq~(\ref{eq:MSD_c1}) and the dotted lines show the MSD of each of the two states. (c) TAMSD for different realization times, the dashed line represents the analytical asymptotic behavior given by Eq.~(\ref{eq:TAMSD_c1}) for $T = 8192$.  
    }
    \label{fig:MSD_3}
\end{figure}

\subsection{Case 3: Two exponential distributions}

For the case where both $\psi_{\pm}(t)$ are exponential, the process $D(t)$ is Markovian and its first moment converges at long times to
\begin{equation}
\langle D(t) \rangle =\langle D_{\text{st}} \rangle= D^+ p^+_{\text{eq}} + D^- p^-_{\text{eq}},
\label{eq:case3_K}
\end{equation}
where $p^{\pm}_{\text{eq}}$ is the probability that a particle is found in state $\pm$,
\begin{equation}
p^{\pm}_{\text{eq}} = \frac{\tau_{\pm}}{\tau_+ + \tau_-}.
\end{equation}
Thus, following Eq.~(\ref{eq:MSD_stat}), the MSD is
\begin{equation}
\langle X^2(t)\rangle = 2 \langle D_{\text{st}} \rangle t^{2H}.
\end{equation}
On the other hand, the TAMSD (Eq.~(\ref{eq:TAMSD_stat}), Fig.~\ref{fig:MSD_3}(c)) is
\begin{equation}
\langle \overline{\delta^2(\Delta)} \rangle =  \frac{\Gamma(2H+1) \Gamma(1-H)}{4^{2H - 1} \Gamma(H+1)}
 \langle D_{\text{st}} \rangle \Delta^{2H}.
\end{equation}
While we have not used Eq.~(\ref{eq:neq_meanK_Laplace}) in this last example, it is possible to proceed in the same fashion as done for the previous two examples with $A=\langle D_{\text{st}} \rangle$ and $B=0$. 

A total of $1000$ trajectories were simulated, all starting in the `$+$' state, with $H = 0.3$, $D^+ = 10$, $D^- = 1$, $\tau_{+} = 15$, and $\tau_{-} = 25$. Figure~\ref{fig:MSD_3}(a) shows a trajectory $X(t)$ alongside with the process $D(t)$. Figure~\ref{fig:MSD_3}(b) shows a comparison between the analytical MSD and the one obtained by numerical simulations. In Figure~\ref{fig:MSD_3}(c), the TAMSD is shown for different realization times, showing that, when $D(t)$ is stationary, there is no dependence on realization time.

% ---- Section ----

\section{Discussion \label{sec:summary}}

In this work, we have studied fractional Brownian motion with fluctuating diffusivity $X(t)$, a process defined in Eq.~(\ref{eq:process}) as a modification of L\'evy's integral representation of the fBm for the case where the generalized diffusion coefficient is a stochastic process $D(t)$. This process, therefore, keeps the temporal correlation while adding an extra layer of complexity encapsulated in the stochastic dynamic of the diffusion coefficient. We derived exact expressions for the MSD (Eq.~(\ref{eq:MSD_laplace})) and TAMSD (Eq.~(\ref{eq:TAMSD_laplace})) in the Laplace domain. For the case when $H = 1/2$, the process in Eq.~(\ref{eq:process}) reduces to a modified Brownian motion with a stochastic diffusion coefficient, for which our expressions reduce to those previously derived for such process. Additionally, when $D(t)$ is a stationary process, its first moment is constant in time, i.e., $\langle D(t) \rangle_{\text{st}} = \langle D \rangle$, and the MSD reduces to that of standard fBm with an effective diffusivity. However, because we employ L\'evy's non-equilibrated representation of the fBm, this system is ultraweakly non-ergodic, where the ensemble averaged MSD has a different prefactor than that of the TAMSD.

To test the capability of our framework, we have considered two-state processes where the diffusion coefficient follows a dichotomous stochastic process. This case is particularly relevant in modeling diffusion in the cytoplasm of live cells \cite{Sabri2020,janczura2021identifying}. In particular, when the sojourn times follow an exponential distribution for one state and a power-law distribution for the other state, has been observed in diverse complex systems \cite{sadegh20141,Sikora2017,
kurilovich2020complex,kurilovich2022non}.  We considered the non-equilibrium dichotomous stochastic process for which the sojourn times for one or both states can have a heavy-tailed distribution. We derived analytical expressions for these processes and found excellent agreement with the numerical simulations.  With our framework, we also find analytical expressions that agree with published numerical simulations \cite{Balcerek2023}. 

All presented numerical simulations start in a defined state. Namely, at $t=0$, the process is in the `$+$' state, and, thus, the MSD at small times has the behavior of that specific state (Figures \ref{fig:MSD_1}(b), \ref{fig:MSD_2}(b) and \ref{fig:MSD_3}(b)), $\langle X^2(t)\rangle = 2 D^+ t^{2H}$. Nevertheless, this dependence is not seen in the TAMSD because this quantity is obtained by averaging over long times, at which the dependence on the initial condition is already lost (Figures \ref{fig:MSD_1}(c), \ref{fig:MSD_2}(c) and \ref{fig:MSD_3}(c)). For systems that can reach equilibrium, such as the Markovian switching described in case 3, it is possible to randomize the initial state so that the system is equilibrated already at $t=0$. Under these conditions, the dependence on the initial condition in the MSD would be eliminated. This type of randomization is highly relevant to experiments where the diffusion process initiated a long time before the measurements \cite{sabri2020elucidating}. Given that the first two cases, when at least one of the states has a heavy-tailed dwell time distribution, do not equilibrate, randomizing the starting state does not eliminate the dependence on initial conditions.

We have considered two-state systems with heavy-tailed dwell time distributions that asymptotically converge toward the state with smaller diffusivity. Namely, the `$-$' state has a smaller $\alpha$ exponent than the `$+$' state, or the latter has an exponential distribution. These conditions lead to the emergence of a crossover between two different temporal diffusion regimes in the MSD (Eqs.~(\ref{eq:MSD_c1}) and~(\ref{eq:MSD_c2}), and Figs.~\ref{fig:MSD_1}(b) and ~\ref{fig:MSD_2}(b)). In a similar fashion as that observed for Brownian motion with fluctuating diffusivities \cite{Miyaguchi2016}, if the state with smaller $\alpha$ exponent has higher $D$, such crossover is not observed and the overall behavior involves a reduction in the MSD at small times followed by a convergence to $2 D^+ t^{2H}$.  

This work opens the way for other possible generalizations, for instance, a situation in which not only the diffusion coefficient is stochastic but also the Hurst exponent. Additionally, the two-state system, here presented as an example, could be generalized to a multi-state system or even to a more complex situation where one of the states involves confinement or transient immobilization.

\begin{acknowledgments}
We thank Ralf Metzler and Micha\l{} Balcerek for useful discussions. This work was supported by the National Science Foundation (NSF) Grant 2102832.
\end{acknowledgments}

\appendix

\section{Laplace transform of the convariance function \label{app:LT_covariance}}
The covariance function can be written as
\begin{multline}
\langle X(t+\Delta)X(t)\rangle = 4 H \int_0^t \langle D(t^{\prime}) \rangle \times \\
\times (t + \Delta - t^{\prime})^{H-1/2} (t - t^{\prime})^{H-1/2} dt^{\prime},
\end{multline}
which, by making the change of variable $u = t - t^{\prime}$ , takes the form of a convolution
\begin{multline}
\langle X(t+\Delta)X(t)\rangle = 4 H \Delta^{2H-1} \int_0^t \langle D(t - u) \rangle\times  \\ 
\times  \left(1 + \frac{u}{\Delta} \right)^{H-1/2} \left(\frac{u}{\Delta} \right)^{H-1/2} du.
\end{multline}
The Laplace transform of a convolution is known and reads
\begin{multline}
\mathcal{L}\left(\langle X(t+\Delta)X(t)\rangle \right) = 4 H \Delta^{2H-1} \langle D(s) \rangle \times \\
\times \int_0^{\infty} e^{-us} \left(1 + \frac{u}{\Delta} \right)^{H-1/2} \left(\frac{u}{\Delta} \right)^{H-1/2} du.
\label{eq:laplace_cov}
\end{multline}
Let us denote the integral in this expression as
\begin{equation}
f(s,\Delta) = \int_0^{\infty} e^{-us} \left(1 - \frac{u}{\Delta} \right)^{H-1/2} \left(\frac{u}{\Delta} \right)^{H-1/2} du.
\end{equation}
Then, by making the change of variable $t = us$, can be rewritten as
\begin{equation}
f(s,\Delta) = \frac{\Gamma(H+1/2)e^{\Delta s / 2}}{\pi^{1/2} \Delta^{H-1} s^{H}} K_H\left( \frac{\Delta s}{2} \right),
\label{eq:integral}
\end{equation}
where $K_{\nu}(z)$ is the modified Bessel function of second kind of order $\nu$ having the following integral representation \cite{Arfken2001}
\begin{multline}
K_{\nu}(z) = \sqrt{\frac{\pi}{2 z}} \frac{e^{-z}}{\Gamma(\nu + 1/2)} \times \\
\times \int_0^{\infty} e^{-t} \left(1 - \frac{t}{2z} \right)^{\nu-1/2} t^{\nu-1/2} d t,
\end{multline}
which can be expressed in terms of the modified Bessel function of first kind $I_{\nu}(z)$ in the following way \cite{Arfken2001}
\begin{equation}
K_{\nu}(z) = \frac{\pi}{2} \frac{I_{-\nu}(z) - I_{\nu}(z)}{\sin (\nu \pi)}.
\end{equation}
This last expression will allow us to find the asymptotic expansion for $s \to 0$. Now, the modified Bessel function of first kind has the following series expansion \cite{Arfken2001}
\begin{equation}
I_{\nu}(z) = \sum_{s = 0}^{\infty} \frac{1}{s! \Gamma(s + \nu + 1)} \left( \frac{z}{2} \right)^{2 s + \nu },
\end{equation}
thus, the modified Bessel function of second kind can be approximated for small $s$ via
\begin{multline}
K_H\left( \frac{\Delta s}{2} \right) = \frac{4^H \pi}{2 \sin(H\pi) (\Delta s )^{H} \Gamma(1 - H) }  \times \\
\times \left[ 1 -  \frac{\Gamma(1-H)}{\Gamma(H+1)} \left( \frac{\Delta s}{4} \right)^{2H} + O\left(s^{2}\right) \right].
\end{multline}
On the other hand, the exponential function in Eq.~(\ref{eq:integral}) can be expanded for $s \to 0$ as
\begin{equation}
\exp \left(\frac{\Delta s}{2} \right) = 1 + \frac{\Delta s}{2} + O(s^2).
\end{equation}
Then, Eq.~(\ref{eq:integral}) can be rewritten as
\begin{multline}
f(s,\Delta) = \frac{\Gamma(2H)}{\Delta^{2H-1} s^{2H}}  \left[ 1 + \frac{\Delta s}{2} - \right. \\ 
\left. -  \frac{\Gamma(1-H)}{\Gamma(H+1)} \left( \frac{\Delta s}{4} \right)^{2H} + O(s^{2H+1}) \right],
\label{eq:int_approx}
\end{multline}
where we have used the following two properties of the Gamma-function \cite{Arfken2001}:
\begin{equation}
\Gamma(x) \Gamma(1-x) = \frac{\pi}{\sin(\pi x)},
\end{equation}
and
\begin{equation}
\Gamma(x+1/2) \Gamma(x) = 2^{1-2x} \pi^{1/2} \Gamma(2x).
\end{equation}
Finally, let us replace Eq.~(\ref{eq:int_approx}) into Eq.~(\ref{eq:laplace_cov}) to find the Laplace transform of the covariance function for $s \to 0$
\begin{multline}
\mathcal{L}\left(\langle X(t+\Delta)X(t)\rangle \right) = 4 H \langle D(s) \rangle \frac{\Gamma(2H)}{ s^{2H}} \times \\
\times \left[ 1 + \frac{\Delta s}{2} -  \frac{\Gamma(1-H)}{\Gamma(H+1)} \left( \frac{\Delta s}{4} \right)^{2H} + O(s^{2H+1}) \right].
\end{multline}

\section{Numerical methods \label{app:Num}}

The numerical methods used to simulate fractional Brownian motion with fluctuating diffusivity follow closely the method presented in Ref.~\cite{Balcerek2023}. Let us start by rewriting the process under consideration, namely, 
\begin{equation}
X(t) = \int_0^t \sqrt{4HD(\tau)} \; (t - \tau)^{H - 1/2} \; dB(\tau).
\label{eq:process_ap}
\end{equation}
Further, we consider the process to start at $t = 0$ and $X(0) = 0$ and to have a realization time $T$. Briefly, the interval $[0,T]$ is divided in $M$ times, $t_i = i \Delta t$, with $i = 1,2,\dots, M$, and $\Delta t = T / M$. From Eq.~(\ref{eq:process_ap}), we known that the process $X(t)$ at each one of the $t = t_i$ is given by 
\begin{equation}
X(t_i) =  \int_0^{t_i} \sqrt{4HD(\tau)} \; (t_i - \tau)^{H - 1/2} \; dB(\tau).
\end{equation}
Next, we divide the interval $[0,t_i]$ into $N$ evenly spaced subintervals $[\tau_j, \tau_{j+1})$ with $j = 0, \dots, N-1$, $\tau_0 = 0$, $\tau_N = t_i$. The above integral can be rewritten as 
\begin{equation}
X(t_i) \approx \sum_{j = 0}^{N-1} \int_{\tau_j}^{\tau_{j+1}} \sqrt{4HD(\tau)} \; (t_i - \tau)^{H - 1/2} \; dB(\tau).
\end{equation}
Each integral can then be Riemann approximated to obtain
\begin{equation}
X(t_i) \approx \sum_{j = 0}^{N-1} \sqrt{4HD(\tau_i)} \; (t_i - \tau_j)^{H - 1/2} \; \xi_j,
\end{equation}
where $\xi_j = B(\tau_{j+i}) - B(\tau_j)$ are i.i.d. Gaussian random numbers with zero mean and variance $\Delta \tau = \tau_{j+1} - \tau_j$. The simulations presented in this work have the following parameters: $\Delta t = 1$ is fixed, $M = T = 10^4$, and $\Delta \tau = 1/50$. Other parameters that were used in the simulations are found in the main text.

\providecommand{\noopsort}[1]{}\providecommand{\singleletter}[1]{#1}%


\begin{thebibliography}{62}%
\makeatletter
\providecommand \@ifxundefined [1]{%
 \@ifx{#1\undefined}
}%
\providecommand \@ifnum [1]{%
 \ifnum #1\expandafter \@firstoftwo
 \else \expandafter \@secondoftwo
 \fi
}%
\providecommand \@ifx [1]{%
 \ifx #1\expandafter \@firstoftwo
 \else \expandafter \@secondoftwo
 \fi
}%
\providecommand \natexlab [1]{#1}%
\providecommand \enquote  [1]{``#1''}%
\providecommand \bibnamefont  [1]{#1}%
\providecommand \bibfnamefont [1]{#1}%
\providecommand \citenamefont [1]{#1}%
\providecommand \href@noop [0]{\@secondoftwo}%
\providecommand \href [0]{\begingroup \@sanitize@url \@href}%
\providecommand \@href[1]{\@@startlink{#1}\@@href}%
\providecommand \@@href[1]{\endgroup#1\@@endlink}%
\providecommand \@sanitize@url [0]{\catcode `\\12\catcode `\$12\catcode
  `\&12\catcode `\#12\catcode `\^12\catcode `\_12\catcode `\%12\relax}%
\providecommand \@@startlink[1]{}%
\providecommand \@@endlink[0]{}%
\providecommand \url  [0]{\begingroup\@sanitize@url \@url }%
\providecommand \@url [1]{\endgroup\@href {#1}{\urlprefix }}%
\providecommand \urlprefix  [0]{URL }%
\providecommand \Eprint [0]{\href }%
\providecommand \doibase [0]{https://doi.org/}%
\providecommand \selectlanguage [0]{\@gobble}%
\providecommand \bibinfo  [0]{\@secondoftwo}%
\providecommand \bibfield  [0]{\@secondoftwo}%
\providecommand \translation [1]{[#1]}%
\providecommand \BibitemOpen [0]{}%
\providecommand \bibitemStop [0]{}%
\providecommand \bibitemNoStop [0]{.\EOS\space}%
\providecommand \EOS [0]{\spacefactor3000\relax}%
\providecommand \BibitemShut  [1]{\csname bibitem#1\endcsname}%
\let\auto@bib@innerbib\@empty
%</preamble>
\bibitem [{\citenamefont {Kindermann}\ \emph {et~al.}(2017)\citenamefont
  {Kindermann}, \citenamefont {Dechant}, \citenamefont {Hohmann}, \citenamefont
  {Lausch}, \citenamefont {Mayer}, \citenamefont {Schmidt}, \citenamefont
  {Lutz},\ and\ \citenamefont {Widera}}]{kindermann2017nonergodic}%
  \BibitemOpen
  \bibfield  {author} {\bibinfo {author} {\bibfnamefont {F.}~\bibnamefont
  {Kindermann}}, \bibinfo {author} {\bibfnamefont {A.}~\bibnamefont {Dechant}},
  \bibinfo {author} {\bibfnamefont {M.}~\bibnamefont {Hohmann}}, \bibinfo
  {author} {\bibfnamefont {T.}~\bibnamefont {Lausch}}, \bibinfo {author}
  {\bibfnamefont {D.}~\bibnamefont {Mayer}}, \bibinfo {author} {\bibfnamefont
  {F.}~\bibnamefont {Schmidt}}, \bibinfo {author} {\bibfnamefont
  {E.}~\bibnamefont {Lutz}},\ and\ \bibinfo {author} {\bibfnamefont
  {A.}~\bibnamefont {Widera}},\ }\bibfield  {title} {\bibinfo {title}
  {Nonergodic diffusion of single atoms in a periodic potential},\ }\href@noop
  {} {\bibfield  {journal} {\bibinfo  {journal} {Nature Physics}\ }\textbf
  {\bibinfo {volume} {13}},\ \bibinfo {pages} {137} (\bibinfo {year}
  {2017})}\BibitemShut {NoStop}%
\bibitem [{\citenamefont {Weigel}\ \emph {et~al.}(2011)\citenamefont {Weigel},
  \citenamefont {Simon}, \citenamefont {Tamkun},\ and\ \citenamefont
  {Krapf}}]{weigel2011ergodic}%
  \BibitemOpen
  \bibfield  {author} {\bibinfo {author} {\bibfnamefont {A.~V.}\ \bibnamefont
  {Weigel}}, \bibinfo {author} {\bibfnamefont {B.}~\bibnamefont {Simon}},
  \bibinfo {author} {\bibfnamefont {M.~M.}\ \bibnamefont {Tamkun}},\ and\
  \bibinfo {author} {\bibfnamefont {D.}~\bibnamefont {Krapf}},\ }\bibfield
  {title} {\bibinfo {title} {Ergodic and nonergodic processes coexist in the
  plasma membrane as observed by single-molecule tracking},\ }\href@noop {}
  {\bibfield  {journal} {\bibinfo  {journal} {Proceedings of the National
  Academy of Sciences}\ }\textbf {\bibinfo {volume} {108}},\ \bibinfo {pages}
  {6438} (\bibinfo {year} {2011})}\BibitemShut {NoStop}%
\bibitem [{\citenamefont {Barkai}\ \emph {et~al.}(2012)\citenamefont {Barkai},
  \citenamefont {Garini},\ and\ \citenamefont {Metzler}}]{barkai2012strange}%
  \BibitemOpen
  \bibfield  {author} {\bibinfo {author} {\bibfnamefont {E.}~\bibnamefont
  {Barkai}}, \bibinfo {author} {\bibfnamefont {Y.}~\bibnamefont {Garini}},\
  and\ \bibinfo {author} {\bibfnamefont {R.}~\bibnamefont {Metzler}},\
  }\bibfield  {title} {\bibinfo {title} {Strange kinetics of single molecules
  in living cells},\ }\href@noop {} {\bibfield  {journal} {\bibinfo  {journal}
  {Physics Today}\ }\textbf {\bibinfo {volume} {65}},\ \bibinfo {pages} {29}
  (\bibinfo {year} {2012})}\BibitemShut {NoStop}%
\bibitem [{\citenamefont {Manzo}\ \emph {et~al.}(2015)\citenamefont {Manzo},
  \citenamefont {Torreno-Pina}, \citenamefont {Massignan}, \citenamefont
  {Lapeyre~Jr}, \citenamefont {Lewenstein},\ and\ \citenamefont
  {Garcia~Parajo}}]{manzo2015weak}%
  \BibitemOpen
  \bibfield  {author} {\bibinfo {author} {\bibfnamefont {C.}~\bibnamefont
  {Manzo}}, \bibinfo {author} {\bibfnamefont {J.~A.}\ \bibnamefont
  {Torreno-Pina}}, \bibinfo {author} {\bibfnamefont {P.}~\bibnamefont
  {Massignan}}, \bibinfo {author} {\bibfnamefont {G.~J.}\ \bibnamefont
  {Lapeyre~Jr}}, \bibinfo {author} {\bibfnamefont {M.}~\bibnamefont
  {Lewenstein}},\ and\ \bibinfo {author} {\bibfnamefont {M.~F.}\ \bibnamefont
  {Garcia~Parajo}},\ }\bibfield  {title} {\bibinfo {title} {Weak ergodicity
  breaking of receptor motion in living cells stemming from random
  diffusivity},\ }\href@noop {} {\bibfield  {journal} {\bibinfo  {journal}
  {Physical Review X}\ }\textbf {\bibinfo {volume} {5}},\ \bibinfo {pages}
  {011021} (\bibinfo {year} {2015})}\BibitemShut {NoStop}%
\bibitem [{\citenamefont {Krapf}\ and\ \citenamefont
  {Metzler}(2019)}]{krapf2019strange}%
  \BibitemOpen
  \bibfield  {author} {\bibinfo {author} {\bibfnamefont {D.}~\bibnamefont
  {Krapf}}\ and\ \bibinfo {author} {\bibfnamefont {R.}~\bibnamefont
  {Metzler}},\ }\bibfield  {title} {\bibinfo {title} {Strange interfacial
  molecular dynamics},\ }\href@noop {} {\bibfield  {journal} {\bibinfo
  {journal} {Physics Today}\ }\textbf {\bibinfo {volume} {72(9)}},\ \bibinfo
  {pages} {48} (\bibinfo {year} {2019})}\BibitemShut {NoStop}%
\bibitem [{\citenamefont {Sabri}\ \emph
  {et~al.}(2020{\natexlab{a}})\citenamefont {Sabri}, \citenamefont {Xu},
  \citenamefont {Krapf},\ and\ \citenamefont {Weiss}}]{sabri2020elucidating}%
  \BibitemOpen
  \bibfield  {author} {\bibinfo {author} {\bibfnamefont {A.}~\bibnamefont
  {Sabri}}, \bibinfo {author} {\bibfnamefont {X.}~\bibnamefont {Xu}}, \bibinfo
  {author} {\bibfnamefont {D.}~\bibnamefont {Krapf}},\ and\ \bibinfo {author}
  {\bibfnamefont {M.}~\bibnamefont {Weiss}},\ }\bibfield  {title} {\bibinfo
  {title} {Elucidating the origin of heterogeneous anomalous diffusion in the
  cytoplasm of mammalian cells},\ }\href@noop {} {\bibfield  {journal}
  {\bibinfo  {journal} {Physical Review Letters}\ }\textbf {\bibinfo {volume}
  {125}},\ \bibinfo {pages} {058101} (\bibinfo {year}
  {2020}{\natexlab{a}})}\BibitemShut {NoStop}%
\bibitem [{\citenamefont {Vilk}\ \emph
  {et~al.}(2022{\natexlab{a}})\citenamefont {Vilk}, \citenamefont {Aghion},
  \citenamefont {Avgar}, \citenamefont {Beta}, \citenamefont {Nagel},
  \citenamefont {Sabri}, \citenamefont {Sarfati}, \citenamefont {Schwartz},
  \citenamefont {Weiss}, \citenamefont {Krapf} \emph
  {et~al.}}]{vilk2022unravelling}%
  \BibitemOpen
  \bibfield  {author} {\bibinfo {author} {\bibfnamefont {O.}~\bibnamefont
  {Vilk}}, \bibinfo {author} {\bibfnamefont {E.}~\bibnamefont {Aghion}},
  \bibinfo {author} {\bibfnamefont {T.}~\bibnamefont {Avgar}}, \bibinfo
  {author} {\bibfnamefont {C.}~\bibnamefont {Beta}}, \bibinfo {author}
  {\bibfnamefont {O.}~\bibnamefont {Nagel}}, \bibinfo {author} {\bibfnamefont
  {A.}~\bibnamefont {Sabri}}, \bibinfo {author} {\bibfnamefont
  {R.}~\bibnamefont {Sarfati}}, \bibinfo {author} {\bibfnamefont {D.~K.}\
  \bibnamefont {Schwartz}}, \bibinfo {author} {\bibfnamefont {M.}~\bibnamefont
  {Weiss}}, \bibinfo {author} {\bibfnamefont {D.}~\bibnamefont {Krapf}}, \emph
  {et~al.},\ }\bibfield  {title} {\bibinfo {title} {Unravelling the origins of
  anomalous diffusion: from molecules to migrating storks},\ }\href@noop {}
  {\bibfield  {journal} {\bibinfo  {journal} {Physical Review Research}\
  }\textbf {\bibinfo {volume} {4}},\ \bibinfo {pages} {033055} (\bibinfo {year}
  {2022}{\natexlab{a}})}\BibitemShut {NoStop}%
\bibitem [{\citenamefont {Vilk}\ \emph
  {et~al.}(2022{\natexlab{b}})\citenamefont {Vilk}, \citenamefont {Orchan},
  \citenamefont {Charter}, \citenamefont {Ganot}, \citenamefont {Toledo},
  \citenamefont {Nathan},\ and\ \citenamefont {Assaf}}]{vilk2022ergodicity}%
  \BibitemOpen
  \bibfield  {author} {\bibinfo {author} {\bibfnamefont {O.}~\bibnamefont
  {Vilk}}, \bibinfo {author} {\bibfnamefont {Y.}~\bibnamefont {Orchan}},
  \bibinfo {author} {\bibfnamefont {M.}~\bibnamefont {Charter}}, \bibinfo
  {author} {\bibfnamefont {N.}~\bibnamefont {Ganot}}, \bibinfo {author}
  {\bibfnamefont {S.}~\bibnamefont {Toledo}}, \bibinfo {author} {\bibfnamefont
  {R.}~\bibnamefont {Nathan}},\ and\ \bibinfo {author} {\bibfnamefont
  {M.}~\bibnamefont {Assaf}},\ }\bibfield  {title} {\bibinfo {title}
  {Ergodicity breaking in area-restricted search of avian predators},\
  }\href@noop {} {\bibfield  {journal} {\bibinfo  {journal} {Physical Review
  X}\ }\textbf {\bibinfo {volume} {12}},\ \bibinfo {pages} {031005} (\bibinfo
  {year} {2022}{\natexlab{b}})}\BibitemShut {NoStop}%
\bibitem [{\citenamefont {Bouchaud}(2005)}]{bouchaud2005subtle}%
  \BibitemOpen
  \bibfield  {author} {\bibinfo {author} {\bibfnamefont {J.-P.}\ \bibnamefont
  {Bouchaud}},\ }\bibfield  {title} {\bibinfo {title} {The subtle nature of
  financial random walks},\ }\href@noop {} {\bibfield  {journal} {\bibinfo
  {journal} {Chaos: An Interdisciplinary Journal of Nonlinear Science}\
  }\textbf {\bibinfo {volume} {15}},\ \bibinfo {pages} {026104} (\bibinfo
  {year} {2005})}\BibitemShut {NoStop}%
\bibitem [{\citenamefont {Scalas}(2006)}]{scalas2006application}%
  \BibitemOpen
  \bibfield  {author} {\bibinfo {author} {\bibfnamefont {E.}~\bibnamefont
  {Scalas}},\ }\bibfield  {title} {\bibinfo {title} {The application of
  continuous-time random walks in finance and economics},\ }\href@noop {}
  {\bibfield  {journal} {\bibinfo  {journal} {Physica A: Statistical Mechanics
  and its Applications}\ }\textbf {\bibinfo {volume} {362}},\ \bibinfo {pages}
  {225} (\bibinfo {year} {2006})}\BibitemShut {NoStop}%
\bibitem [{\citenamefont {Plerou}\ \emph {et~al.}(2000)\citenamefont {Plerou},
  \citenamefont {Gopikrishnan}, \citenamefont {Amaral}, \citenamefont
  {Gabaix},\ and\ \citenamefont {Stanley}}]{plerou2000economic}%
  \BibitemOpen
  \bibfield  {author} {\bibinfo {author} {\bibfnamefont {V.}~\bibnamefont
  {Plerou}}, \bibinfo {author} {\bibfnamefont {P.}~\bibnamefont
  {Gopikrishnan}}, \bibinfo {author} {\bibfnamefont {L.~A.~N.}\ \bibnamefont
  {Amaral}}, \bibinfo {author} {\bibfnamefont {X.}~\bibnamefont {Gabaix}},\
  and\ \bibinfo {author} {\bibfnamefont {H.~E.}\ \bibnamefont {Stanley}},\
  }\bibfield  {title} {\bibinfo {title} {Economic fluctuations and anomalous
  diffusion},\ }\href@noop {} {\bibfield  {journal} {\bibinfo  {journal}
  {Physical Review E}\ }\textbf {\bibinfo {volume} {62}},\ \bibinfo {pages}
  {R3023} (\bibinfo {year} {2000})}\BibitemShut {NoStop}%
\bibitem [{\citenamefont {H{\"o}fling}\ and\ \citenamefont
  {Franosch}(2013)}]{hofling2013anomalous}%
  \BibitemOpen
  \bibfield  {author} {\bibinfo {author} {\bibfnamefont {F.}~\bibnamefont
  {H{\"o}fling}}\ and\ \bibinfo {author} {\bibfnamefont {T.}~\bibnamefont
  {Franosch}},\ }\bibfield  {title} {\bibinfo {title} {Anomalous transport in
  the crowded world of biological cells},\ }\href@noop {} {\bibfield  {journal}
  {\bibinfo  {journal} {Reports on Progress in Physics}\ }\textbf {\bibinfo
  {volume} {76}},\ \bibinfo {pages} {046602} (\bibinfo {year}
  {2013})}\BibitemShut {NoStop}%
\bibitem [{\citenamefont {Metzler}\ \emph {et~al.}(2014)\citenamefont
  {Metzler}, \citenamefont {Jeon}, \citenamefont {Cherstvy},\ and\
  \citenamefont {Barkai}}]{metzler2014anomalous}%
  \BibitemOpen
  \bibfield  {author} {\bibinfo {author} {\bibfnamefont {R.}~\bibnamefont
  {Metzler}}, \bibinfo {author} {\bibfnamefont {J.-H.}\ \bibnamefont {Jeon}},
  \bibinfo {author} {\bibfnamefont {A.~G.}\ \bibnamefont {Cherstvy}},\ and\
  \bibinfo {author} {\bibfnamefont {E.}~\bibnamefont {Barkai}},\ }\bibfield
  {title} {\bibinfo {title} {Anomalous diffusion models and their properties:
  non-stationarity, non-ergodicity, and ageing at the centenary of single
  particle tracking},\ }\href@noop {} {\bibfield  {journal} {\bibinfo
  {journal} {Physical Chemistry Chemical Physics}\ }\textbf {\bibinfo {volume}
  {16}},\ \bibinfo {pages} {24128} (\bibinfo {year} {2014})}\BibitemShut
  {NoStop}%
\bibitem [{\citenamefont {Manzo}\ and\ \citenamefont
  {Garcia-Parajo}(2015)}]{manzo2015review}%
  \BibitemOpen
  \bibfield  {author} {\bibinfo {author} {\bibfnamefont {C.}~\bibnamefont
  {Manzo}}\ and\ \bibinfo {author} {\bibfnamefont {M.~F.}\ \bibnamefont
  {Garcia-Parajo}},\ }\bibfield  {title} {\bibinfo {title} {A review of
  progress in single particle tracking: from methods to biophysical insights},\
  }\href@noop {} {\bibfield  {journal} {\bibinfo  {journal} {Reports on
  Progress in Physics}\ }\textbf {\bibinfo {volume} {78}},\ \bibinfo {pages}
  {124601} (\bibinfo {year} {2015})}\BibitemShut {NoStop}%
\bibitem [{\citenamefont {Krapf}(2015)}]{krapf2015mechanisms}%
  \BibitemOpen
  \bibfield  {author} {\bibinfo {author} {\bibfnamefont {D.}~\bibnamefont
  {Krapf}},\ }\bibfield  {title} {\bibinfo {title} {Mechanisms underlying
  anomalous diffusion in the plasma membrane},\ }\href@noop {} {\bibfield
  {journal} {\bibinfo  {journal} {Current Topics in Membranes}\ }\textbf
  {\bibinfo {volume} {75}},\ \bibinfo {pages} {167} (\bibinfo {year}
  {2015})}\BibitemShut {NoStop}%
\bibitem [{\citenamefont {Shen}\ \emph {et~al.}(2017)\citenamefont {Shen},
  \citenamefont {Tauzin}, \citenamefont {Baiyasi}, \citenamefont {Wang},
  \citenamefont {Moringo}, \citenamefont {Shuang},\ and\ \citenamefont
  {Landes}}]{shen2017single}%
  \BibitemOpen
  \bibfield  {author} {\bibinfo {author} {\bibfnamefont {H.}~\bibnamefont
  {Shen}}, \bibinfo {author} {\bibfnamefont {L.~J.}\ \bibnamefont {Tauzin}},
  \bibinfo {author} {\bibfnamefont {R.}~\bibnamefont {Baiyasi}}, \bibinfo
  {author} {\bibfnamefont {W.}~\bibnamefont {Wang}}, \bibinfo {author}
  {\bibfnamefont {N.}~\bibnamefont {Moringo}}, \bibinfo {author} {\bibfnamefont
  {B.}~\bibnamefont {Shuang}},\ and\ \bibinfo {author} {\bibfnamefont {C.~F.}\
  \bibnamefont {Landes}},\ }\bibfield  {title} {\bibinfo {title} {Single
  particle tracking: from theory to biophysical applications},\ }\href@noop {}
  {\bibfield  {journal} {\bibinfo  {journal} {Chemical reviews}\ }\textbf
  {\bibinfo {volume} {117}},\ \bibinfo {pages} {7331} (\bibinfo {year}
  {2017})}\BibitemShut {NoStop}%
\bibitem [{\citenamefont {Mu{\~n}oz-Gil}\ \emph {et~al.}(2021)\citenamefont
  {Mu{\~n}oz-Gil}, \citenamefont {Volpe}, \citenamefont {Garcia-March},
  \citenamefont {Aghion}, \citenamefont {Argun}, \citenamefont {Hong},
  \citenamefont {Bland}, \citenamefont {Bo}, \citenamefont {Conejero},
  \citenamefont {Firbas} \emph {et~al.}}]{munoz2021objective}%
  \BibitemOpen
  \bibfield  {author} {\bibinfo {author} {\bibfnamefont {G.}~\bibnamefont
  {Mu{\~n}oz-Gil}}, \bibinfo {author} {\bibfnamefont {G.}~\bibnamefont
  {Volpe}}, \bibinfo {author} {\bibfnamefont {M.~A.}\ \bibnamefont
  {Garcia-March}}, \bibinfo {author} {\bibfnamefont {E.}~\bibnamefont
  {Aghion}}, \bibinfo {author} {\bibfnamefont {A.}~\bibnamefont {Argun}},
  \bibinfo {author} {\bibfnamefont {C.~B.}\ \bibnamefont {Hong}}, \bibinfo
  {author} {\bibfnamefont {T.}~\bibnamefont {Bland}}, \bibinfo {author}
  {\bibfnamefont {S.}~\bibnamefont {Bo}}, \bibinfo {author} {\bibfnamefont
  {J.~A.}\ \bibnamefont {Conejero}}, \bibinfo {author} {\bibfnamefont
  {N.}~\bibnamefont {Firbas}}, \emph {et~al.},\ }\bibfield  {title} {\bibinfo
  {title} {Objective comparison of methods to decode anomalous diffusion},\
  }\href@noop {} {\bibfield  {journal} {\bibinfo  {journal} {Nature
  Communications}\ }\textbf {\bibinfo {volume} {12}},\ \bibinfo {pages} {6253}
  (\bibinfo {year} {2021})}\BibitemShut {NoStop}%
\bibitem [{\citenamefont {Kolmogorov}(1940)}]{kolmogorov1940wienersche}%
  \BibitemOpen
  \bibfield  {author} {\bibinfo {author} {\bibfnamefont {A.~N.}\ \bibnamefont
  {Kolmogorov}},\ }\bibfield  {title} {\bibinfo {title} {Wienersche spiralen
  und einige andere interessante {K}urven in {H}ilbertscen {R}aum, {C.R.}
  (doklady)},\ }\href@noop {} {\bibfield  {journal} {\bibinfo  {journal} {Acad.
  Sci. URSS (NS)}\ }\textbf {\bibinfo {volume} {26}},\ \bibinfo {pages} {115}
  (\bibinfo {year} {1940})}\BibitemShut {NoStop}%
\bibitem [{\citenamefont {Mandelbrot}\ and\ \citenamefont
  {Van~Ness}(1968)}]{Mandelbrot1968}%
  \BibitemOpen
  \bibfield  {author} {\bibinfo {author} {\bibfnamefont {B.~B.}\ \bibnamefont
  {Mandelbrot}}\ and\ \bibinfo {author} {\bibfnamefont {J.~W.}\ \bibnamefont
  {Van~Ness}},\ }\bibfield  {title} {\bibinfo {title} {Fractional {B}rownian
  motions, fractional noises and applications},\ }\href
  {https://doi.org/10.1137/1010093} {\bibfield  {journal} {\bibinfo  {journal}
  {SIAM Review}\ }\textbf {\bibinfo {volume} {10}},\ \bibinfo {pages} {422}
  (\bibinfo {year} {1968})}\BibitemShut {NoStop}%
\bibitem [{\citenamefont {Szymanski}\ and\ \citenamefont
  {Weiss}(2009)}]{Szymanski2009}%
  \BibitemOpen
  \bibfield  {author} {\bibinfo {author} {\bibfnamefont {J.}~\bibnamefont
  {Szymanski}}\ and\ \bibinfo {author} {\bibfnamefont {M.}~\bibnamefont
  {Weiss}},\ }\bibfield  {title} {\bibinfo {title} {Elucidating the origin of
  anomalous diffusion in crowded fluids},\ }\href
  {https://doi.org/10.1103/PhysRevLett.103.038102} {\bibfield  {journal}
  {\bibinfo  {journal} {Physical Review Letters}\ }\textbf {\bibinfo {volume}
  {103}},\ \bibinfo {pages} {038102} (\bibinfo {year} {2009})}\BibitemShut
  {NoStop}%
\bibitem [{\citenamefont {Guigas}\ \emph {et~al.}(2007)\citenamefont {Guigas},
  \citenamefont {Kalla},\ and\ \citenamefont {Weiss}}]{guigas2007probing}%
  \BibitemOpen
  \bibfield  {author} {\bibinfo {author} {\bibfnamefont {G.}~\bibnamefont
  {Guigas}}, \bibinfo {author} {\bibfnamefont {C.}~\bibnamefont {Kalla}},\ and\
  \bibinfo {author} {\bibfnamefont {M.}~\bibnamefont {Weiss}},\ }\bibfield
  {title} {\bibinfo {title} {Probing the nanoscale viscoelasticity of
  intracellular fluids in living cells},\ }\href@noop {} {\bibfield  {journal}
  {\bibinfo  {journal} {Biophysical Journal}\ }\textbf {\bibinfo {volume}
  {93}},\ \bibinfo {pages} {316} (\bibinfo {year} {2007})}\BibitemShut
  {NoStop}%
\bibitem [{\citenamefont {Magdziarz}\ \emph {et~al.}(2009)\citenamefont
  {Magdziarz}, \citenamefont {Weron}, \citenamefont {Burnecki},\ and\
  \citenamefont {Klafter}}]{Magdziarz2009}%
  \BibitemOpen
  \bibfield  {author} {\bibinfo {author} {\bibfnamefont {M.}~\bibnamefont
  {Magdziarz}}, \bibinfo {author} {\bibfnamefont {A.}~\bibnamefont {Weron}},
  \bibinfo {author} {\bibfnamefont {K.}~\bibnamefont {Burnecki}},\ and\
  \bibinfo {author} {\bibfnamefont {J.}~\bibnamefont {Klafter}},\ }\bibfield
  {title} {\bibinfo {title} {Fractional {B}rownian motion versus the
  continuous-time random walk: A simple test for subdiffusive dynamics},\
  }\href {https://doi.org/10.1103/PhysRevLett.103.180602} {\bibfield  {journal}
  {\bibinfo  {journal} {Physical Review Letters}\ }\textbf {\bibinfo {volume}
  {103}},\ \bibinfo {pages} {180602} (\bibinfo {year} {2009})}\BibitemShut
  {NoStop}%
\bibitem [{\citenamefont {Jeon}\ \emph {et~al.}(2011)\citenamefont {Jeon},
  \citenamefont {Tejedor}, \citenamefont {Burov}, \citenamefont {Barkai},
  \citenamefont {Selhuber-Unkel}, \citenamefont {Berg-S{\o}rensen},
  \citenamefont {Oddershede},\ and\ \citenamefont {Metzler}}]{jeon2011vivo}%
  \BibitemOpen
  \bibfield  {author} {\bibinfo {author} {\bibfnamefont {J.-H.}\ \bibnamefont
  {Jeon}}, \bibinfo {author} {\bibfnamefont {V.}~\bibnamefont {Tejedor}},
  \bibinfo {author} {\bibfnamefont {S.}~\bibnamefont {Burov}}, \bibinfo
  {author} {\bibfnamefont {E.}~\bibnamefont {Barkai}}, \bibinfo {author}
  {\bibfnamefont {C.}~\bibnamefont {Selhuber-Unkel}}, \bibinfo {author}
  {\bibfnamefont {K.}~\bibnamefont {Berg-S{\o}rensen}}, \bibinfo {author}
  {\bibfnamefont {L.}~\bibnamefont {Oddershede}},\ and\ \bibinfo {author}
  {\bibfnamefont {R.}~\bibnamefont {Metzler}},\ }\bibfield  {title} {\bibinfo
  {title} {In vivo anomalous diffusion and weak ergodicity breaking of lipid
  granules},\ }\href@noop {} {\bibfield  {journal} {\bibinfo  {journal}
  {Physical Review Letters}\ }\textbf {\bibinfo {volume} {106}},\ \bibinfo
  {pages} {048103} (\bibinfo {year} {2011})}\BibitemShut {NoStop}%
\bibitem [{\citenamefont {Jeon}\ \emph {et~al.}(2013)\citenamefont {Jeon},
  \citenamefont {Leijnse}, \citenamefont {Oddershede},\ and\ \citenamefont
  {Metzler}}]{jeon2013anomalous}%
  \BibitemOpen
  \bibfield  {author} {\bibinfo {author} {\bibfnamefont {J.-H.}\ \bibnamefont
  {Jeon}}, \bibinfo {author} {\bibfnamefont {N.}~\bibnamefont {Leijnse}},
  \bibinfo {author} {\bibfnamefont {L.~B.}\ \bibnamefont {Oddershede}},\ and\
  \bibinfo {author} {\bibfnamefont {R.}~\bibnamefont {Metzler}},\ }\bibfield
  {title} {\bibinfo {title} {Anomalous diffusion and power-law relaxation of
  the time averaged mean squared displacement in worm-like micellar
  solutions},\ }\href@noop {} {\bibfield  {journal} {\bibinfo  {journal} {New
  Journal of Physics}\ }\textbf {\bibinfo {volume} {15}},\ \bibinfo {pages}
  {045011} (\bibinfo {year} {2013})}\BibitemShut {NoStop}%
\bibitem [{\citenamefont {Sadegh}\ \emph {et~al.}(2017)\citenamefont {Sadegh},
  \citenamefont {Higgins}, \citenamefont {Mannion}, \citenamefont {Tamkun},\
  and\ \citenamefont {Krapf}}]{Sadegh2017}%
  \BibitemOpen
  \bibfield  {author} {\bibinfo {author} {\bibfnamefont {S.}~\bibnamefont
  {Sadegh}}, \bibinfo {author} {\bibfnamefont {J.~L.}\ \bibnamefont {Higgins}},
  \bibinfo {author} {\bibfnamefont {P.~C.}\ \bibnamefont {Mannion}}, \bibinfo
  {author} {\bibfnamefont {M.~M.}\ \bibnamefont {Tamkun}},\ and\ \bibinfo
  {author} {\bibfnamefont {D.}~\bibnamefont {Krapf}},\ }\bibfield  {title}
  {\bibinfo {title} {Plasma membrane is compartmentalized by a self-similar
  cortical actin meshwork},\ }\href {https://doi.org/10.1103/PhysRevX.7.011031}
  {\bibfield  {journal} {\bibinfo  {journal} {Physical Review X}\ }\textbf
  {\bibinfo {volume} {7}},\ \bibinfo {pages} {011031} (\bibinfo {year}
  {2017})}\BibitemShut {NoStop}%
\bibitem [{\citenamefont {Waigh}\ and\ \citenamefont
  {Korabel}(2023)}]{waigh2023heterogeneous}%
  \BibitemOpen
  \bibfield  {author} {\bibinfo {author} {\bibfnamefont {T.~A.}\ \bibnamefont
  {Waigh}}\ and\ \bibinfo {author} {\bibfnamefont {N.}~\bibnamefont
  {Korabel}},\ }\bibfield  {title} {\bibinfo {title} {Heterogeneous anomalous
  transport in cellular and molecular biology},\ }\href@noop {} {\bibfield
  {journal} {\bibinfo  {journal} {Reports on Progress in Physics}\ } (\bibinfo
  {year} {2023})}\BibitemShut {NoStop}%
\bibitem [{\citenamefont {Lanoisel{\'e}e}\ \emph {et~al.}(2018)\citenamefont
  {Lanoisel{\'e}e}, \citenamefont {Moutal},\ and\ \citenamefont
  {Grebenkov}}]{lanoiselee2018diffusion}%
  \BibitemOpen
  \bibfield  {author} {\bibinfo {author} {\bibfnamefont {Y.}~\bibnamefont
  {Lanoisel{\'e}e}}, \bibinfo {author} {\bibfnamefont {N.}~\bibnamefont
  {Moutal}},\ and\ \bibinfo {author} {\bibfnamefont {D.~S.}\ \bibnamefont
  {Grebenkov}},\ }\bibfield  {title} {\bibinfo {title} {Diffusion-limited
  reactions in dynamic heterogeneous media},\ }\href@noop {} {\bibfield
  {journal} {\bibinfo  {journal} {Nature Communications}\ }\textbf {\bibinfo
  {volume} {9}},\ \bibinfo {pages} {4398} (\bibinfo {year} {2018})}\BibitemShut
  {NoStop}%
\bibitem [{\citenamefont {He}\ \emph {et~al.}(2016)\citenamefont {He},
  \citenamefont {Song}, \citenamefont {Su}, \citenamefont {Geng}, \citenamefont
  {Ackerson}, \citenamefont {Peng},\ and\ \citenamefont {Tong}}]{He2016}%
  \BibitemOpen
  \bibfield  {author} {\bibinfo {author} {\bibfnamefont {W.}~\bibnamefont
  {He}}, \bibinfo {author} {\bibfnamefont {H.}~\bibnamefont {Song}}, \bibinfo
  {author} {\bibfnamefont {Y.}~\bibnamefont {Su}}, \bibinfo {author}
  {\bibfnamefont {L.}~\bibnamefont {Geng}}, \bibinfo {author} {\bibfnamefont
  {B.~J.}\ \bibnamefont {Ackerson}}, \bibinfo {author} {\bibfnamefont
  {H.}~\bibnamefont {Peng}},\ and\ \bibinfo {author} {\bibfnamefont
  {P.}~\bibnamefont {Tong}},\ }\bibfield  {title} {\bibinfo {title} {Dynamic
  heterogeneity and non-{G}aussian statistics for acetylcholine receptors on
  live cell membrane},\ }\href@noop {} {\bibfield  {journal} {\bibinfo
  {journal} {Nature Communications}\ }\textbf {\bibinfo {volume} {7}},\
  \bibinfo {pages} {11701} (\bibinfo {year} {2016})}\BibitemShut {NoStop}%
\bibitem [{\citenamefont {Jeon}\ \emph {et~al.}(2016)\citenamefont {Jeon},
  \citenamefont {Javanainen}, \citenamefont {Martinez-Seara}, \citenamefont
  {Metzler},\ and\ \citenamefont {Vattulainen}}]{Jeon2016}%
  \BibitemOpen
  \bibfield  {author} {\bibinfo {author} {\bibfnamefont {J.-H.}\ \bibnamefont
  {Jeon}}, \bibinfo {author} {\bibfnamefont {M.}~\bibnamefont {Javanainen}},
  \bibinfo {author} {\bibfnamefont {H.}~\bibnamefont {Martinez-Seara}},
  \bibinfo {author} {\bibfnamefont {R.}~\bibnamefont {Metzler}},\ and\ \bibinfo
  {author} {\bibfnamefont {I.}~\bibnamefont {Vattulainen}},\ }\bibfield
  {title} {\bibinfo {title} {Protein crowding in lipid bilayers gives rise to
  non-{G}aussian anomalous lateral diffusion of phospholipids and proteins},\
  }\href@noop {} {\bibfield  {journal} {\bibinfo  {journal} {Physical Review
  X}\ }\textbf {\bibinfo {volume} {6}},\ \bibinfo {pages} {021006} (\bibinfo
  {year} {2016})}\BibitemShut {NoStop}%
\bibitem [{\citenamefont {Sikora}\ \emph {et~al.}(2017)\citenamefont {Sikora},
  \citenamefont {Wy{\l}oma{\'n}ska}, \citenamefont {Gajda}, \citenamefont
  {Sol{\'e}}, \citenamefont {Akin}, \citenamefont {Tamkun},\ and\ \citenamefont
  {Krapf}}]{Sikora2017}%
  \BibitemOpen
  \bibfield  {author} {\bibinfo {author} {\bibfnamefont {G.}~\bibnamefont
  {Sikora}}, \bibinfo {author} {\bibfnamefont {A.}~\bibnamefont
  {Wy{\l}oma{\'n}ska}}, \bibinfo {author} {\bibfnamefont {J.}~\bibnamefont
  {Gajda}}, \bibinfo {author} {\bibfnamefont {L.}~\bibnamefont {Sol{\'e}}},
  \bibinfo {author} {\bibfnamefont {E.~J.}\ \bibnamefont {Akin}}, \bibinfo
  {author} {\bibfnamefont {M.~M.}\ \bibnamefont {Tamkun}},\ and\ \bibinfo
  {author} {\bibfnamefont {D.}~\bibnamefont {Krapf}},\ }\bibfield  {title}
  {\bibinfo {title} {Elucidating distinct ion channel populations on the
  surface of hippocampal neurons via single-particle tracking recurrence
  analysis},\ }\href@noop {} {\bibfield  {journal} {\bibinfo  {journal}
  {Physical Review E}\ }\textbf {\bibinfo {volume} {96}},\ \bibinfo {pages}
  {062404} (\bibinfo {year} {2017})}\BibitemShut {NoStop}%
\bibitem [{\citenamefont {Weron}\ \emph {et~al.}(2017)\citenamefont {Weron},
  \citenamefont {Burnecki}, \citenamefont {Akin}, \citenamefont {Sol{\'e}},
  \citenamefont {Balcerek}, \citenamefont {Tamkun},\ and\ \citenamefont
  {Krapf}}]{Weron2017}%
  \BibitemOpen
  \bibfield  {author} {\bibinfo {author} {\bibfnamefont {A.}~\bibnamefont
  {Weron}}, \bibinfo {author} {\bibfnamefont {K.}~\bibnamefont {Burnecki}},
  \bibinfo {author} {\bibfnamefont {E.~J.}\ \bibnamefont {Akin}}, \bibinfo
  {author} {\bibfnamefont {L.}~\bibnamefont {Sol{\'e}}}, \bibinfo {author}
  {\bibfnamefont {M.}~\bibnamefont {Balcerek}}, \bibinfo {author}
  {\bibfnamefont {M.~M.}\ \bibnamefont {Tamkun}},\ and\ \bibinfo {author}
  {\bibfnamefont {D.}~\bibnamefont {Krapf}},\ }\bibfield  {title} {\bibinfo
  {title} {Ergodicity breaking on the neuronal surface emerges from random
  switching between diffusive states},\ }\href@noop {} {\bibfield  {journal}
  {\bibinfo  {journal} {Scientific Reports}\ }\textbf {\bibinfo {volume} {7}},\
  \bibinfo {pages} {5404} (\bibinfo {year} {2017})}\BibitemShut {NoStop}%
\bibitem [{\citenamefont {Han}\ \emph {et~al.}(2020)\citenamefont {Han},
  \citenamefont {Korabel}, \citenamefont {Chen}, \citenamefont {Johnston},
  \citenamefont {Gavrilova}, \citenamefont {Allan}, \citenamefont {Fedotov},\
  and\ \citenamefont {Waigh}}]{Han2020}%
  \BibitemOpen
  \bibfield  {author} {\bibinfo {author} {\bibfnamefont {D.}~\bibnamefont
  {Han}}, \bibinfo {author} {\bibfnamefont {N.}~\bibnamefont {Korabel}},
  \bibinfo {author} {\bibfnamefont {R.}~\bibnamefont {Chen}}, \bibinfo {author}
  {\bibfnamefont {M.}~\bibnamefont {Johnston}}, \bibinfo {author}
  {\bibfnamefont {A.}~\bibnamefont {Gavrilova}}, \bibinfo {author}
  {\bibfnamefont {V.~J.}\ \bibnamefont {Allan}}, \bibinfo {author}
  {\bibfnamefont {S.}~\bibnamefont {Fedotov}},\ and\ \bibinfo {author}
  {\bibfnamefont {T.~A.}\ \bibnamefont {Waigh}},\ }\bibfield  {title} {\bibinfo
  {title} {Deciphering anomalous heterogeneous intracellular transport with
  neural networks},\ }\href@noop {} {\bibfield  {journal} {\bibinfo  {journal}
  {eLife}\ }\textbf {\bibinfo {volume} {9}},\ \bibinfo {pages} {e52224}
  (\bibinfo {year} {2020})}\BibitemShut {NoStop}%
\bibitem [{\citenamefont {Fedotov}\ and\ \citenamefont
  {Han}(2023)}]{Fedotov2023}%
  \BibitemOpen
  \bibfield  {author} {\bibinfo {author} {\bibfnamefont {S.}~\bibnamefont
  {Fedotov}}\ and\ \bibinfo {author} {\bibfnamefont {D.}~\bibnamefont {Han}},\
  }\bibfield  {title} {\bibinfo {title} {Population heterogeneity in the
  fractional master equation, ensemble self-reinforcement, and strong memory
  effects},\ }\href@noop {} {\bibfield  {journal} {\bibinfo  {journal}
  {Physical Review E}\ }\textbf {\bibinfo {volume} {107}},\ \bibinfo {pages}
  {034115} (\bibinfo {year} {2023})}\BibitemShut {NoStop}%
\bibitem [{\citenamefont {Loverdo}\ \emph {et~al.}(2009)\citenamefont
  {Loverdo}, \citenamefont {B\'enichou}, \citenamefont {Voituriez},
  \citenamefont {Biebricher}, \citenamefont {Bonnet},\ and\ \citenamefont
  {Desbiolles}}]{Loverdo2009}%
  \BibitemOpen
  \bibfield  {author} {\bibinfo {author} {\bibfnamefont {C.}~\bibnamefont
  {Loverdo}}, \bibinfo {author} {\bibfnamefont {O.}~\bibnamefont {B\'enichou}},
  \bibinfo {author} {\bibfnamefont {R.}~\bibnamefont {Voituriez}}, \bibinfo
  {author} {\bibfnamefont {A.}~\bibnamefont {Biebricher}}, \bibinfo {author}
  {\bibfnamefont {I.}~\bibnamefont {Bonnet}},\ and\ \bibinfo {author}
  {\bibfnamefont {P.}~\bibnamefont {Desbiolles}},\ }\bibfield  {title}
  {\bibinfo {title} {Quantifying hopping and jumping in facilitated diffusion
  of {DNA}-binding proteins},\ }\href
  {https://doi.org/10.1103/PhysRevLett.102.188101} {\bibfield  {journal}
  {\bibinfo  {journal} {Physical Review Letters}\ }\textbf {\bibinfo {volume}
  {102}},\ \bibinfo {pages} {188101} (\bibinfo {year} {2009})}\BibitemShut
  {NoStop}%
\bibitem [{\citenamefont {Mu{\~n}oz}\ \emph {et~al.}(2023)\citenamefont
  {Mu{\~n}oz}, \citenamefont {Bachimanchi}, \citenamefont {Pineda},
  \citenamefont {Midtvedt}, \citenamefont {Lewenstein}, \citenamefont
  {Metzler}, \citenamefont {Krapf}, \citenamefont {Volpe},\ and\ \citenamefont
  {Manzo}}]{Munoz2023}%
  \BibitemOpen
  \bibfield  {author} {\bibinfo {author} {\bibfnamefont {G.}~\bibnamefont
  {Mu{\~n}oz}}, \bibinfo {author} {\bibfnamefont {H.}~\bibnamefont
  {Bachimanchi}}, \bibinfo {author} {\bibfnamefont {J.}~\bibnamefont {Pineda}},
  \bibinfo {author} {\bibfnamefont {B.}~\bibnamefont {Midtvedt}}, \bibinfo
  {author} {\bibfnamefont {M.}~\bibnamefont {Lewenstein}}, \bibinfo {author}
  {\bibfnamefont {R.}~\bibnamefont {Metzler}}, \bibinfo {author} {\bibfnamefont
  {D.}~\bibnamefont {Krapf}}, \bibinfo {author} {\bibfnamefont
  {G.}~\bibnamefont {Volpe}},\ and\ \bibinfo {author} {\bibfnamefont
  {C.}~\bibnamefont {Manzo}},\ }\bibfield  {title} {\bibinfo {title}
  {Quantitative evaluation of methods to analyze motion changes in
  single-particle experiments},\ }\href@noop {} {\bibfield  {journal} {\bibinfo
   {journal} {arXiv preprint arXiv:2311.18100}\ } (\bibinfo {year}
  {2023})}\BibitemShut {NoStop}%
\bibitem [{\citenamefont {Bronstein}\ \emph {et~al.}(2009)\citenamefont
  {Bronstein}, \citenamefont {Israel}, \citenamefont {Kepten}, \citenamefont
  {Mai}, \citenamefont {Shav-Tal}, \citenamefont {Barkai},\ and\ \citenamefont
  {Garini}}]{Bronstein2009}%
  \BibitemOpen
  \bibfield  {author} {\bibinfo {author} {\bibfnamefont {I.}~\bibnamefont
  {Bronstein}}, \bibinfo {author} {\bibfnamefont {Y.}~\bibnamefont {Israel}},
  \bibinfo {author} {\bibfnamefont {E.}~\bibnamefont {Kepten}}, \bibinfo
  {author} {\bibfnamefont {S.}~\bibnamefont {Mai}}, \bibinfo {author}
  {\bibfnamefont {Y.}~\bibnamefont {Shav-Tal}}, \bibinfo {author}
  {\bibfnamefont {E.}~\bibnamefont {Barkai}},\ and\ \bibinfo {author}
  {\bibfnamefont {Y.}~\bibnamefont {Garini}},\ }\bibfield  {title} {\bibinfo
  {title} {Transient anomalous diffusion of telomeres in the nucleus of
  mammalian cells},\ }\href@noop {} {\bibfield  {journal} {\bibinfo  {journal}
  {Physical review letters}\ }\textbf {\bibinfo {volume} {103}},\ \bibinfo
  {pages} {018102} (\bibinfo {year} {2009})}\BibitemShut {NoStop}%
\bibitem [{\citenamefont {Sabri}\ \emph
  {et~al.}(2020{\natexlab{b}})\citenamefont {Sabri}, \citenamefont {Xu},
  \citenamefont {Krapf},\ and\ \citenamefont {Weiss}}]{Sabri2020}%
  \BibitemOpen
  \bibfield  {author} {\bibinfo {author} {\bibfnamefont {A.}~\bibnamefont
  {Sabri}}, \bibinfo {author} {\bibfnamefont {X.}~\bibnamefont {Xu}}, \bibinfo
  {author} {\bibfnamefont {D.}~\bibnamefont {Krapf}},\ and\ \bibinfo {author}
  {\bibfnamefont {M.}~\bibnamefont {Weiss}},\ }\bibfield  {title} {\bibinfo
  {title} {Elucidating the origin of heterogeneous anomalous diffusion in the
  cytoplasm of mammalian cells},\ }\href@noop {} {\bibfield  {journal}
  {\bibinfo  {journal} {Physical Review Letters}\ }\textbf {\bibinfo {volume}
  {125}},\ \bibinfo {pages} {058101} (\bibinfo {year}
  {2020}{\natexlab{b}})}\BibitemShut {NoStop}%
\bibitem [{\citenamefont {Chubynsky}\ and\ \citenamefont
  {Slater}(2014)}]{chubynsky2014diffusing}%
  \BibitemOpen
  \bibfield  {author} {\bibinfo {author} {\bibfnamefont {M.~V.}\ \bibnamefont
  {Chubynsky}}\ and\ \bibinfo {author} {\bibfnamefont {G.~W.}\ \bibnamefont
  {Slater}},\ }\bibfield  {title} {\bibinfo {title} {Diffusing diffusivity: a
  model for anomalous, yet brownian, diffusion},\ }\href@noop {} {\bibfield
  {journal} {\bibinfo  {journal} {Physical Review Letters}\ }\textbf {\bibinfo
  {volume} {113}},\ \bibinfo {pages} {098302} (\bibinfo {year}
  {2014})}\BibitemShut {NoStop}%
\bibitem [{\citenamefont {Chechkin}\ \emph {et~al.}(2017)\citenamefont
  {Chechkin}, \citenamefont {Seno}, \citenamefont {Metzler},\ and\
  \citenamefont {Sokolov}}]{chechkin2017brownian}%
  \BibitemOpen
  \bibfield  {author} {\bibinfo {author} {\bibfnamefont {A.~V.}\ \bibnamefont
  {Chechkin}}, \bibinfo {author} {\bibfnamefont {F.}~\bibnamefont {Seno}},
  \bibinfo {author} {\bibfnamefont {R.}~\bibnamefont {Metzler}},\ and\ \bibinfo
  {author} {\bibfnamefont {I.~M.}\ \bibnamefont {Sokolov}},\ }\bibfield
  {title} {\bibinfo {title} {Brownian yet non-gaussian diffusion: from
  superstatistics to subordination of diffusing diffusivities},\ }\href@noop {}
  {\bibfield  {journal} {\bibinfo  {journal} {Physical Review X}\ }\textbf
  {\bibinfo {volume} {7}},\ \bibinfo {pages} {021002} (\bibinfo {year}
  {2017})}\BibitemShut {NoStop}%
\bibitem [{\citenamefont {Massignan}\ \emph {et~al.}(2014)\citenamefont
  {Massignan}, \citenamefont {Manzo}, \citenamefont {Torreno-Pina},
  \citenamefont {Garc{\'\i}a-Parajo}, \citenamefont {Lewenstein},\ and\
  \citenamefont {Lapeyre~Jr}}]{massignan2014nonergodic}%
  \BibitemOpen
  \bibfield  {author} {\bibinfo {author} {\bibfnamefont {P.}~\bibnamefont
  {Massignan}}, \bibinfo {author} {\bibfnamefont {C.}~\bibnamefont {Manzo}},
  \bibinfo {author} {\bibfnamefont {J.~A.}\ \bibnamefont {Torreno-Pina}},
  \bibinfo {author} {\bibfnamefont {M.~F.}\ \bibnamefont {Garc{\'\i}a-Parajo}},
  \bibinfo {author} {\bibfnamefont {M.}~\bibnamefont {Lewenstein}},\ and\
  \bibinfo {author} {\bibfnamefont {G.}~\bibnamefont {Lapeyre~Jr}},\ }\bibfield
   {title} {\bibinfo {title} {Nonergodic subdiffusion from brownian motion in
  an inhomogeneous medium},\ }\href@noop {} {\bibfield  {journal} {\bibinfo
  {journal} {Physical Review Letters}\ }\textbf {\bibinfo {volume} {112}},\
  \bibinfo {pages} {150603} (\bibinfo {year} {2014})}\BibitemShut {NoStop}%
\bibitem [{\citenamefont {Wang}\ \emph
  {et~al.}(2020{\natexlab{a}})\citenamefont {Wang}, \citenamefont {Seno},
  \citenamefont {Sokolov}, \citenamefont {Chechkin},\ and\ \citenamefont
  {Metzler}}]{Wang2020}%
  \BibitemOpen
  \bibfield  {author} {\bibinfo {author} {\bibfnamefont {W.}~\bibnamefont
  {Wang}}, \bibinfo {author} {\bibfnamefont {F.}~\bibnamefont {Seno}}, \bibinfo
  {author} {\bibfnamefont {I.~M.}\ \bibnamefont {Sokolov}}, \bibinfo {author}
  {\bibfnamefont {A.~V.}\ \bibnamefont {Chechkin}},\ and\ \bibinfo {author}
  {\bibfnamefont {R.}~\bibnamefont {Metzler}},\ }\bibfield  {title} {\bibinfo
  {title} {Unexpected crossovers in correlated random-diffusivity processes},\
  }\href {https://doi.org/10.1088/1367-2630/aba390} {\bibfield  {journal}
  {\bibinfo  {journal} {New Journal of Physics}\ }\textbf {\bibinfo {volume}
  {22}},\ \bibinfo {pages} {083041} (\bibinfo {year}
  {2020}{\natexlab{a}})}\BibitemShut {NoStop}%
\bibitem [{\citenamefont {Wang}\ \emph
  {et~al.}(2020{\natexlab{b}})\citenamefont {Wang}, \citenamefont {Cherstvy},
  \citenamefont {Chechkin}, \citenamefont {Thapa}, \citenamefont {Seno},
  \citenamefont {Liu},\ and\ \citenamefont {Metzler}}]{Wang2020_1}%
  \BibitemOpen
  \bibfield  {author} {\bibinfo {author} {\bibfnamefont {W.}~\bibnamefont
  {Wang}}, \bibinfo {author} {\bibfnamefont {A.~G.}\ \bibnamefont {Cherstvy}},
  \bibinfo {author} {\bibfnamefont {A.~V.}\ \bibnamefont {Chechkin}}, \bibinfo
  {author} {\bibfnamefont {S.}~\bibnamefont {Thapa}}, \bibinfo {author}
  {\bibfnamefont {F.}~\bibnamefont {Seno}}, \bibinfo {author} {\bibfnamefont
  {X.}~\bibnamefont {Liu}},\ and\ \bibinfo {author} {\bibfnamefont
  {R.}~\bibnamefont {Metzler}},\ }\bibfield  {title} {\bibinfo {title}
  {Fractional {B}rownian motion with random diffusivity: emerging residual
  nonergodicity below the correlation time},\ }\href
  {https://doi.org/10.1088/1751-8121/aba467} {\bibfield  {journal} {\bibinfo
  {journal} {Journal of Physics A: Mathematical and Theoretical}\ }\textbf
  {\bibinfo {volume} {53}},\ \bibinfo {pages} {474001} (\bibinfo {year}
  {2020}{\natexlab{b}})}\BibitemShut {NoStop}%
\bibitem [{\citenamefont {Fox}\ \emph {et~al.}(2021)\citenamefont {Fox},
  \citenamefont {Barkai},\ and\ \citenamefont {Krapf}}]{Fox2021}%
  \BibitemOpen
  \bibfield  {author} {\bibinfo {author} {\bibfnamefont {Z.~R.}\ \bibnamefont
  {Fox}}, \bibinfo {author} {\bibfnamefont {E.}~\bibnamefont {Barkai}},\ and\
  \bibinfo {author} {\bibfnamefont {D.}~\bibnamefont {Krapf}},\ }\bibfield
  {title} {\bibinfo {title} {Aging power spectrum of membrane protein transport
  and other subordinated random walks},\ }\href
  {https://doi.org/10.1038/s41467-021-26465-8} {\bibfield  {journal} {\bibinfo
  {journal} {Nature Communications}\ }\textbf {\bibinfo {volume} {12}},\
  \bibinfo {pages} {6162} (\bibinfo {year} {2021})}\BibitemShut {NoStop}%
\bibitem [{\citenamefont {Szarek}\ \emph {et~al.}(2022)\citenamefont {Szarek},
  \citenamefont {Jabłoński}, \citenamefont {Krapf},\ and\ \citenamefont
  {Wyłomańska}}]{Szarek2022}%
  \BibitemOpen
  \bibfield  {author} {\bibinfo {author} {\bibfnamefont {D.}~\bibnamefont
  {Szarek}}, \bibinfo {author} {\bibfnamefont {I.}~\bibnamefont {Jabłoński}},
  \bibinfo {author} {\bibfnamefont {D.}~\bibnamefont {Krapf}},\ and\ \bibinfo
  {author} {\bibfnamefont {A.}~\bibnamefont {Wyłomańska}},\ }\bibfield
  {title} {\bibinfo {title} {{Multifractional {B}rownian motion
  characterization based on {H}urst exponent estimation and statistical
  learning}},\ }\href {https://doi.org/10.1063/5.0093836} {\bibfield  {journal}
  {\bibinfo  {journal} {Chaos: An Interdisciplinary Journal of Nonlinear
  Science}\ }\textbf {\bibinfo {volume} {32}},\ \bibinfo {pages} {083148}
  (\bibinfo {year} {2022})}\BibitemShut {NoStop}%
\bibitem [{\citenamefont {Balcerek}\ \emph {et~al.}(2022)\citenamefont
  {Balcerek}, \citenamefont {Burnecki}, \citenamefont {Thapa}, \citenamefont
  {Wyłomańska},\ and\ \citenamefont {Chechkin}}]{Balcerek2022}%
  \BibitemOpen
  \bibfield  {author} {\bibinfo {author} {\bibfnamefont {M.}~\bibnamefont
  {Balcerek}}, \bibinfo {author} {\bibfnamefont {K.}~\bibnamefont {Burnecki}},
  \bibinfo {author} {\bibfnamefont {S.}~\bibnamefont {Thapa}}, \bibinfo
  {author} {\bibfnamefont {A.}~\bibnamefont {Wyłomańska}},\ and\ \bibinfo
  {author} {\bibfnamefont {A.}~\bibnamefont {Chechkin}},\ }\bibfield  {title}
  {\bibinfo {title} {{Fractional {B}rownian motion with random {H}urst
  exponent: Accelerating diffusion and persistence transitions}},\ }\href
  {https://doi.org/10.1063/5.0101913} {\bibfield  {journal} {\bibinfo
  {journal} {Chaos: An Interdisciplinary Journal of Nonlinear Science}\
  }\textbf {\bibinfo {volume} {32}},\ \bibinfo {pages} {093114} (\bibinfo
  {year} {2022})}\BibitemShut {NoStop}%
\bibitem [{\citenamefont {Balcerek}\ \emph {et~al.}(2023)\citenamefont
  {Balcerek}, \citenamefont {Wyłomańska}, \citenamefont {Burnecki},
  \citenamefont {Metzler},\ and\ \citenamefont {Krapf}}]{Balcerek2023}%
  \BibitemOpen
  \bibfield  {author} {\bibinfo {author} {\bibfnamefont {M.}~\bibnamefont
  {Balcerek}}, \bibinfo {author} {\bibfnamefont {A.}~\bibnamefont
  {Wyłomańska}}, \bibinfo {author} {\bibfnamefont {K.}~\bibnamefont
  {Burnecki}}, \bibinfo {author} {\bibfnamefont {R.}~\bibnamefont {Metzler}},\
  and\ \bibinfo {author} {\bibfnamefont {D.}~\bibnamefont {Krapf}},\ }\bibfield
   {title} {\bibinfo {title} {Modelling intermittent anomalous diffusion with
  switching fractional {B}rownian motion},\ }\href
  {https://doi.org/10.1088/1367-2630/ad00d7} {\bibfield  {journal} {\bibinfo
  {journal} {New Journal of Physics}\ }\textbf {\bibinfo {volume} {25}},\
  \bibinfo {pages} {103031} (\bibinfo {year} {2023})}\BibitemShut {NoStop}%
\bibitem [{\citenamefont {Wang}\ \emph {et~al.}(2023)\citenamefont {Wang},
  \citenamefont {Balcerek}, \citenamefont {Burnecki}, \citenamefont {Chechkin},
  \citenamefont {Janu{\v{s}}onis}, \citenamefont {{\'S}lezak}, \citenamefont
  {Vojta}, \citenamefont {Wy{\l}oma{\'n}ska},\ and\ \citenamefont
  {Metzler}}]{Wang2023}%
  \BibitemOpen
  \bibfield  {author} {\bibinfo {author} {\bibfnamefont {W.}~\bibnamefont
  {Wang}}, \bibinfo {author} {\bibfnamefont {M.}~\bibnamefont {Balcerek}},
  \bibinfo {author} {\bibfnamefont {K.}~\bibnamefont {Burnecki}}, \bibinfo
  {author} {\bibfnamefont {A.~V.}\ \bibnamefont {Chechkin}}, \bibinfo {author}
  {\bibfnamefont {S.}~\bibnamefont {Janu{\v{s}}onis}}, \bibinfo {author}
  {\bibfnamefont {J.}~\bibnamefont {{\'S}lezak}}, \bibinfo {author}
  {\bibfnamefont {T.}~\bibnamefont {Vojta}}, \bibinfo {author} {\bibfnamefont
  {A.}~\bibnamefont {Wy{\l}oma{\'n}ska}},\ and\ \bibinfo {author}
  {\bibfnamefont {R.}~\bibnamefont {Metzler}},\ }\bibfield  {title} {\bibinfo
  {title} {Memory-multi-fractional {B}rownian motion with continuous
  correlations},\ }\href@noop {} {\bibfield  {journal} {\bibinfo  {journal}
  {Physical Review Research}\ }\textbf {\bibinfo {volume} {5}},\ \bibinfo
  {pages} {L032025} (\bibinfo {year} {2023})}\BibitemShut {NoStop}%
\bibitem [{\citenamefont {{\'S}lezak}\ and\ \citenamefont
  {Metzler}(2023)}]{Slezak2023}%
  \BibitemOpen
  \bibfield  {author} {\bibinfo {author} {\bibfnamefont {J.}~\bibnamefont
  {{\'S}lezak}}\ and\ \bibinfo {author} {\bibfnamefont {R.}~\bibnamefont
  {Metzler}},\ }\bibfield  {title} {\bibinfo {title} {Minimal model of
  diffusion with time changing {H}urst exponent},\ }\href@noop {} {\bibfield
  {journal} {\bibinfo  {journal} {Journal of Physics A: Mathematical and
  Theoretical}\ }\textbf {\bibinfo {volume} {56}},\ \bibinfo {pages} {35LT01}
  (\bibinfo {year} {2023})}\BibitemShut {NoStop}%
\bibitem [{\citenamefont {L{\'e}vy}(1953{\natexlab{a}})}]{Levy1953}%
  \BibitemOpen
  \bibfield  {author} {\bibinfo {author} {\bibfnamefont {P.}~\bibnamefont
  {L{\'e}vy}},\ }\href {https://books.google.com/books?id=t2VVAAAAMAAJ} {\emph
  {\bibinfo {title} {Random Functions: General Theory with Special Reference to
  Laplacian Random Functions}}},\ University of California publications in
  statistics\ (\bibinfo  {publisher} {University of California Press},\
  \bibinfo {year} {1953})\BibitemShut {NoStop}%
\bibitem [{\citenamefont {Janczura}\ \emph {et~al.}(2021)\citenamefont
  {Janczura}, \citenamefont {Balcerek}, \citenamefont {Burnecki}, \citenamefont
  {Sabri}, \citenamefont {Weiss},\ and\ \citenamefont
  {Krapf}}]{janczura2021identifying}%
  \BibitemOpen
  \bibfield  {author} {\bibinfo {author} {\bibfnamefont {J.}~\bibnamefont
  {Janczura}}, \bibinfo {author} {\bibfnamefont {M.}~\bibnamefont {Balcerek}},
  \bibinfo {author} {\bibfnamefont {K.}~\bibnamefont {Burnecki}}, \bibinfo
  {author} {\bibfnamefont {A.}~\bibnamefont {Sabri}}, \bibinfo {author}
  {\bibfnamefont {M.}~\bibnamefont {Weiss}},\ and\ \bibinfo {author}
  {\bibfnamefont {D.}~\bibnamefont {Krapf}},\ }\bibfield  {title} {\bibinfo
  {title} {Identifying heterogeneous diffusion states in the cytoplasm by a
  hidden {M}arkov model},\ }\href@noop {} {\bibfield  {journal} {\bibinfo
  {journal} {New Journal of Physics}\ }\textbf {\bibinfo {volume} {23}},\
  \bibinfo {pages} {053018} (\bibinfo {year} {2021})}\BibitemShut {NoStop}%
\bibitem [{\citenamefont {Miyaguchi}\ \emph {et~al.}(2016)\citenamefont
  {Miyaguchi}, \citenamefont {Akimoto},\ and\ \citenamefont
  {Yamamoto}}]{Miyaguchi2016}%
  \BibitemOpen
  \bibfield  {author} {\bibinfo {author} {\bibfnamefont {T.}~\bibnamefont
  {Miyaguchi}}, \bibinfo {author} {\bibfnamefont {T.}~\bibnamefont {Akimoto}},\
  and\ \bibinfo {author} {\bibfnamefont {E.}~\bibnamefont {Yamamoto}},\
  }\bibfield  {title} {\bibinfo {title} {Langevin equation with fluctuating
  diffusivity: A two-state model},\ }\href
  {https://doi.org/10.1103/PhysRevE.94.012109} {\bibfield  {journal} {\bibinfo
  {journal} {Physical Review E}\ }\textbf {\bibinfo {volume} {94}},\ \bibinfo
  {pages} {012109} (\bibinfo {year} {2016})}\BibitemShut {NoStop}%
\bibitem [{\citenamefont {Dieball}\ \emph {et~al.}(2022)\citenamefont
  {Dieball}, \citenamefont {Krapf}, \citenamefont {Weiss},\ and\ \citenamefont
  {Godec}}]{dieball2022scattering}%
  \BibitemOpen
  \bibfield  {author} {\bibinfo {author} {\bibfnamefont {C.}~\bibnamefont
  {Dieball}}, \bibinfo {author} {\bibfnamefont {D.}~\bibnamefont {Krapf}},
  \bibinfo {author} {\bibfnamefont {M.}~\bibnamefont {Weiss}},\ and\ \bibinfo
  {author} {\bibfnamefont {A.}~\bibnamefont {Godec}},\ }\bibfield  {title}
  {\bibinfo {title} {Scattering fingerprints of two-state dynamics},\
  }\href@noop {} {\bibfield  {journal} {\bibinfo  {journal} {New Journal of
  Physics}\ }\textbf {\bibinfo {volume} {24}},\ \bibinfo {pages} {023004}
  (\bibinfo {year} {2022})}\BibitemShut {NoStop}%
\bibitem [{\citenamefont {Weigel}\ \emph {et~al.}(2013)\citenamefont {Weigel},
  \citenamefont {Tamkun},\ and\ \citenamefont {Krapf}}]{weigel2013quantifying}%
  \BibitemOpen
  \bibfield  {author} {\bibinfo {author} {\bibfnamefont {A.~V.}\ \bibnamefont
  {Weigel}}, \bibinfo {author} {\bibfnamefont {M.~M.}\ \bibnamefont {Tamkun}},\
  and\ \bibinfo {author} {\bibfnamefont {D.}~\bibnamefont {Krapf}},\ }\bibfield
   {title} {\bibinfo {title} {Quantifying the dynamic interactions between a
  clathrin-coated pit and cargo molecules},\ }\href@noop {} {\bibfield
  {journal} {\bibinfo  {journal} {Proceedings of the National Academy of
  Sciences}\ }\textbf {\bibinfo {volume} {110}},\ \bibinfo {pages} {E4591}
  (\bibinfo {year} {2013})}\BibitemShut {NoStop}%
\bibitem [{\citenamefont {Tabei}\ \emph {et~al.}(2013)\citenamefont {Tabei},
  \citenamefont {Burov}, \citenamefont {Kim}, \citenamefont {Kuznetsov},
  \citenamefont {Huynh}, \citenamefont {Jureller}, \citenamefont {Philipson},
  \citenamefont {Dinner},\ and\ \citenamefont
  {Scherer}}]{tabei2013intracellular}%
  \BibitemOpen
  \bibfield  {author} {\bibinfo {author} {\bibfnamefont {S.~A.}\ \bibnamefont
  {Tabei}}, \bibinfo {author} {\bibfnamefont {S.}~\bibnamefont {Burov}},
  \bibinfo {author} {\bibfnamefont {H.~Y.}\ \bibnamefont {Kim}}, \bibinfo
  {author} {\bibfnamefont {A.}~\bibnamefont {Kuznetsov}}, \bibinfo {author}
  {\bibfnamefont {T.}~\bibnamefont {Huynh}}, \bibinfo {author} {\bibfnamefont
  {J.}~\bibnamefont {Jureller}}, \bibinfo {author} {\bibfnamefont {L.~H.}\
  \bibnamefont {Philipson}}, \bibinfo {author} {\bibfnamefont {A.~R.}\
  \bibnamefont {Dinner}},\ and\ \bibinfo {author} {\bibfnamefont {N.~F.}\
  \bibnamefont {Scherer}},\ }\bibfield  {title} {\bibinfo {title}
  {Intracellular transport of insulin granules is a subordinated random walk},\
  }\href@noop {} {\bibfield  {journal} {\bibinfo  {journal} {Proceedings of the
  National Academy of Sciences}\ }\textbf {\bibinfo {volume} {110}},\ \bibinfo
  {pages} {4911} (\bibinfo {year} {2013})}\BibitemShut {NoStop}%
\bibitem [{\citenamefont {Hu}\ \emph {et~al.}(2016)\citenamefont {Hu},
  \citenamefont {Hong}, \citenamefont {Dean~Smith}, \citenamefont {Neusius},
  \citenamefont {Cheng},\ and\ \citenamefont {Smith}}]{hu2016dynamics}%
  \BibitemOpen
  \bibfield  {author} {\bibinfo {author} {\bibfnamefont {X.}~\bibnamefont
  {Hu}}, \bibinfo {author} {\bibfnamefont {L.}~\bibnamefont {Hong}}, \bibinfo
  {author} {\bibfnamefont {M.}~\bibnamefont {Dean~Smith}}, \bibinfo {author}
  {\bibfnamefont {T.}~\bibnamefont {Neusius}}, \bibinfo {author} {\bibfnamefont
  {X.}~\bibnamefont {Cheng}},\ and\ \bibinfo {author} {\bibfnamefont {J.~C.}\
  \bibnamefont {Smith}},\ }\bibfield  {title} {\bibinfo {title} {The dynamics
  of single protein molecules is non-equilibrium and self-similar over thirteen
  decades in time},\ }\href@noop {} {\bibfield  {journal} {\bibinfo  {journal}
  {Nature Physics}\ }\textbf {\bibinfo {volume} {12}},\ \bibinfo {pages} {171}
  (\bibinfo {year} {2016})}\BibitemShut {NoStop}%
\bibitem [{\citenamefont {Yang}\ \emph {et~al.}(2003)\citenamefont {Yang},
  \citenamefont {Luo}, \citenamefont {Karnchanaphanurach}, \citenamefont
  {Louie}, \citenamefont {Rech}, \citenamefont {Cova}, \citenamefont {Xun},\
  and\ \citenamefont {Xie}}]{yang2003protein}%
  \BibitemOpen
  \bibfield  {author} {\bibinfo {author} {\bibfnamefont {H.}~\bibnamefont
  {Yang}}, \bibinfo {author} {\bibfnamefont {G.}~\bibnamefont {Luo}}, \bibinfo
  {author} {\bibfnamefont {P.}~\bibnamefont {Karnchanaphanurach}}, \bibinfo
  {author} {\bibfnamefont {T.-M.}\ \bibnamefont {Louie}}, \bibinfo {author}
  {\bibfnamefont {I.}~\bibnamefont {Rech}}, \bibinfo {author} {\bibfnamefont
  {S.}~\bibnamefont {Cova}}, \bibinfo {author} {\bibfnamefont {L.}~\bibnamefont
  {Xun}},\ and\ \bibinfo {author} {\bibfnamefont {X.~S.}\ \bibnamefont {Xie}},\
  }\bibfield  {title} {\bibinfo {title} {Protein conformational dynamics probed
  by single-molecule electron transfer},\ }\href@noop {} {\bibfield  {journal}
  {\bibinfo  {journal} {Science}\ }\textbf {\bibinfo {volume} {302}},\ \bibinfo
  {pages} {262} (\bibinfo {year} {2003})}\BibitemShut {NoStop}%
\bibitem [{\citenamefont {L{\'e}vy}(1953{\natexlab{b}})}]{levy1953random}%
  \BibitemOpen
  \bibfield  {author} {\bibinfo {author} {\bibfnamefont {P.}~\bibnamefont
  {L{\'e}vy}},\ }\href@noop {} {\emph {\bibinfo {title} {Random functions:
  general theory with special reference to {L}aplacian random functions}}}\
  (\bibinfo  {publisher} {University of California Press},\ \bibinfo {year}
  {1953})\BibitemShut {NoStop}%
\bibitem [{\citenamefont {Marinucci}\ and\ \citenamefont
  {Robinson}(1999)}]{Marinucci1999}%
  \BibitemOpen
  \bibfield  {author} {\bibinfo {author} {\bibfnamefont {D.}~\bibnamefont
  {Marinucci}}\ and\ \bibinfo {author} {\bibfnamefont {P.}~\bibnamefont
  {Robinson}},\ }\bibfield  {title} {\bibinfo {title} {Alternative forms of
  fractional {B}rownian motion},\ }\href
  {https://doi.org/https://doi.org/10.1016/S0378-3758(98)00245-6} {\bibfield
  {journal} {\bibinfo  {journal} {Journal of Statistical Planning and
  Inference}\ }\textbf {\bibinfo {volume} {80}},\ \bibinfo {pages} {111}
  (\bibinfo {year} {1999})}\BibitemShut {NoStop}%
\bibitem [{\citenamefont {Sadegh}\ \emph {et~al.}(2014)\citenamefont {Sadegh},
  \citenamefont {Barkai},\ and\ \citenamefont {Krapf}}]{sadegh20141}%
  \BibitemOpen
  \bibfield  {author} {\bibinfo {author} {\bibfnamefont {S.}~\bibnamefont
  {Sadegh}}, \bibinfo {author} {\bibfnamefont {E.}~\bibnamefont {Barkai}},\
  and\ \bibinfo {author} {\bibfnamefont {D.}~\bibnamefont {Krapf}},\ }\bibfield
   {title} {\bibinfo {title} {1/f noise for intermittent quantum dots exhibits
  non-stationarity and critical exponents},\ }\href@noop {} {\bibfield
  {journal} {\bibinfo  {journal} {New Journal of Physics}\ }\textbf {\bibinfo
  {volume} {16}},\ \bibinfo {pages} {113054} (\bibinfo {year}
  {2014})}\BibitemShut {NoStop}%
\bibitem [{\citenamefont {Kurilovich}\ \emph {et~al.}(2020)\citenamefont
  {Kurilovich}, \citenamefont {Mantsevich}, \citenamefont {Stevenson},
  \citenamefont {Chechkin},\ and\ \citenamefont
  {Palyulin}}]{kurilovich2020complex}%
  \BibitemOpen
  \bibfield  {author} {\bibinfo {author} {\bibfnamefont {A.~A.}\ \bibnamefont
  {Kurilovich}}, \bibinfo {author} {\bibfnamefont {V.~N.}\ \bibnamefont
  {Mantsevich}}, \bibinfo {author} {\bibfnamefont {K.~J.}\ \bibnamefont
  {Stevenson}}, \bibinfo {author} {\bibfnamefont {A.~V.}\ \bibnamefont
  {Chechkin}},\ and\ \bibinfo {author} {\bibfnamefont {V.~V.}\ \bibnamefont
  {Palyulin}},\ }\bibfield  {title} {\bibinfo {title} {Complex diffusion-based
  kinetics of photoluminescence in semiconductor nanoplatelets},\ }\href@noop
  {} {\bibfield  {journal} {\bibinfo  {journal} {Physical Chemistry Chemical
  Physics}\ }\textbf {\bibinfo {volume} {22}},\ \bibinfo {pages} {24686}
  (\bibinfo {year} {2020})}\BibitemShut {NoStop}%
\bibitem [{\citenamefont {Kurilovich}\ \emph {et~al.}(2022)\citenamefont
  {Kurilovich}, \citenamefont {Mantsevich}, \citenamefont {Mardoukhi},
  \citenamefont {Stevenson}, \citenamefont {Chechkin},\ and\ \citenamefont
  {Palyulin}}]{kurilovich2022non}%
  \BibitemOpen
  \bibfield  {author} {\bibinfo {author} {\bibfnamefont {A.~A.}\ \bibnamefont
  {Kurilovich}}, \bibinfo {author} {\bibfnamefont {V.~N.}\ \bibnamefont
  {Mantsevich}}, \bibinfo {author} {\bibfnamefont {Y.}~\bibnamefont
  {Mardoukhi}}, \bibinfo {author} {\bibfnamefont {K.~J.}\ \bibnamefont
  {Stevenson}}, \bibinfo {author} {\bibfnamefont {A.~V.}\ \bibnamefont
  {Chechkin}},\ and\ \bibinfo {author} {\bibfnamefont {V.~V.}\ \bibnamefont
  {Palyulin}},\ }\bibfield  {title} {\bibinfo {title} {Non-{M}arkovian
  diffusion of excitons in layered perovskites and transition metal
  dichalcogenides},\ }\href@noop {} {\bibfield  {journal} {\bibinfo  {journal}
  {Physical Chemistry Chemical Physics}\ }\textbf {\bibinfo {volume} {24}},\
  \bibinfo {pages} {13941} (\bibinfo {year} {2022})}\BibitemShut {NoStop}%
\bibitem [{\citenamefont {Arfken}\ and\ \citenamefont
  {Weber}(2001)}]{Arfken2001}%
  \BibitemOpen
  \bibfield  {author} {\bibinfo {author} {\bibfnamefont {G.}~\bibnamefont
  {Arfken}}\ and\ \bibinfo {author} {\bibfnamefont {H.}~\bibnamefont {Weber}},\
  }\href {https://books.google.com/books?id=QXzwAAAAMAAJ} {\emph {\bibinfo
  {title} {Mathematical Methods for Physicists}}},\ Mathematical Methods for
  Physicists\ (\bibinfo  {publisher} {Harcourt/Academic Press},\ \bibinfo
  {year} {2001})\BibitemShut {NoStop}%
\end{thebibliography}
\end{document}